\documentclass[aps,prd,a4paper,superscriptaddress,nofootinbib,10pt, two column]{revtex4-1}

\usepackage{amsmath,amssymb,graphicx,multirow,dcolumn,bm,latexsym,soul,nicefrac}
\usepackage{epstopdf}
\usepackage{mathrsfs}
\usepackage{acronym}
\usepackage[colorlinks,linkcolor=blue,citecolor=blue,urlcolor=blue ]{hyperref}
\usepackage{ulem}
\usepackage{subcaption}
\usepackage{float}
\usepackage{xcolor}

\usepackage[labelformat=simple,labelsep=period,skip=3pt]{caption}
\newcounter{RomanNumber}

\allowdisplaybreaks

\newcommand{\HBU}{Department of Physics, Hebei University, Baoding, 071002, China. \\
Hebei Key Laboratory of High-precision Computation and Application of Quantum Field Theory, Baoding, 071002, China. \\
Hebei Research Center of the Basic Discipline for Computational Physics, Baoding, 071002, China.}
\newcommand{\SYU}{ School of Physics and Astronomy, Sun Yat-sen University (Zhuhai Campus), Zhuhai 519082, China. \\
MOE Key Laboratory of TianQin Mission, TianQin Research Center for
Gravitational Physics, Frontiers Science Center for TianQin, \\
 Gravitational Wave Research Center of CNSA, Sun Yat-sen University (Zhuhai Campus), Zhuhai 519082, China.}

\begin{document}

\title{Probing the  Extreme-Mass-Ratio Inspirals Population Constraints with TianQin}

\author{Hui-Min Fan}
\affiliation{\HBU}

\author{Xiang-Yu Lyu}
\affiliation{\SYU}

\author{Jian-dong Zhang}
\affiliation{\SYU}

\author{Yi-Ming Hu}
\email{huyiming@mail.sysu.edu.cn}
\affiliation{\SYU}

\author{Rong-Jia Yang}
\email{yangrongjia@tsinghua.org.cn}
\affiliation{\HBU}

\author{Tai-Fu Feng}
\affiliation{\HBU}

\date{\today}

\begin{abstract}
Extreme-mass-ratio inspirals (EMRIs), consisting of a massive black hole and a stellar compact object, are one of the most important sources for space-borne gravitational wave detectors like TianQin. Their population study can be used to constrain astrophysical models that interpret the EMRI formation mechanisms. In this paper, as an initial attempt, we employ a parametrization method to describe the EMRI population model in the loss cone formation channel. This approach, however, can be extended to other models such as the accretion disc driven formation channel. We present the phenomenological characteristic of the MBH mass, spin, and redshift distributions. Then, we investigate the posterior distribution of the hyper-parameters that describe this population model. The optimistic results show that TianQin could recover almost all the posterior of the hyper-parameters within $1\sigma$ confidence interval. The hyper-parameters $\alpha_1, \alpha_2, b$, which describe the MBH mass distribution, could be measured with an accuracy of $46.4\%$, $12.6\%$, and $3\%$, respectively. The hyper-parameters $\mu_z$, and $\sigma_z$, which describe the redshift distribution, could be measured with an accuracy of $15.4\%$ and $21.1\%$. With this estimation accuracy, the EMRI population characteristics can be effectively demonstrated, potentially serving as evidence for EMRI formation in the future studies. Furthermore, with an increasing number of detectable events, the parameter estimation for the hyper-parameters will improve and the confidence intervals will be narrowed. 

\end{abstract}
\maketitle

%%%%%%%%%%%%%%%%% BODY OF PAPER %%%%%%%%%%%%%%%%%%

\section{Introduction}

The population properties of black holes and neutron stars are being extensively analyzed using data from the LIGO-Virgo Gravitational-wave Transient Catalog \cite{Talbot:2017yur, LIGOScientific:2018jsj, LIGOScientific:2020kqk,KAGRA:2021duu}. By studying these detected events, existing models of compact binary formation are being validated.  Unlike the ground-based interferometric detectors \cite{LIGOScientific:2016vbw}, the space-borne gravitational wave (GW) detectors, such as LISA and TianQin \cite{amaro2017laser, TianQin:2015yph}, have not yet started operating. These space-borne GW detectors, designed with longer armlengths, are sensitive to heavier sources, like those involving massive black holes (MBHs) or even the low-frequency inspiral phase of stellar-origin compact binaries \cite{Baker:2019nia, TianQin:2020hid}. Given the wide array of formation channels proposed for the target sources of the space-borne GW detectors \cite{alexander2017emris,Pan:2021oob,Naoz:2022rru,Bortolas:2019sif}, the population study can become very important if one wishes to understand the exact formation and evolution of those GW sources. In this paper, we'd like to evaluate how well the theoretically predicted population models of those GW sources can be constrained, and thereby make forecasts for future astrophysical analysis.

The space-borne GW detector, TianQin \cite{TianQin:2015yph}, is designed to detect GW signals in the frequency band $10^{-4}-1 \rm Hz$ \cite{Fan:2022wio, Ye:2023aeo}. Its target sources include Galactic ultra-compact binaries \cite{Hu:2018yqb,Brown:2020uvh,Kremer:2017xrg,Korol:2017qcx}; coalescing massive black holes \cite{Feng:2019wgq,Ruan:2021fxq,Shuman:2021ruh,Katz:2019qlu}; the mergers of intermediate-mass black holes \cite{Fragione:2022ams, Torres-Orjuela:2023hfd}; the low-frequency inspirals of stellar-mass black holes \cite{Klein:2022rbf,Buscicchio:2021dph,Ewing:2020brd,Toubiana:2020cqv}; the extreme-mass ratio inspirals \cite{Zhang:2022xuq,Wardell:2021fyy,Lynch:2021ogr,Isoyama:2021jjd,Vazquez-Aceves:2022dgi}; and  the stochastic GW backgrounds \cite{Renzini:2022alw,LISACosmologyWorkingGroup:2022kbp,Boileau:2020rpg}. Among these sources,  EMRIs are significant for allowing for testing the gravitational theories in the strong field regime \cite{Amaro-Seoane:2007osp, Amaro-Seoane:2012lgq}, and for checking the validity of the black hole no-hair theorem \cite{Zi:2021pdp, Shi:2019hqa}. Beyond the values from individual EMRI systems, the statistical properties of the set of EMRI detectable events are highly valuable in constraining population models. This allows us to make inferences about the EMRI physics, gain a better understanding of their origins, and identify candidate host galaxies  to infer the history of cosmic expansion \cite{ LIGOScientific:2020kqk, Gair:2008bx, Gair:2010yu, Gair:2010bx}.

So far, a number of studies have been performed exploring the science prospects of various sources with TianQin   \cite{Huang:2020rjf,Wang:2019ryf,Liu:2020eko,Fan:2020zhy,Liang:2021bde}. For EMRIs, it is expected that TianQin will detect tens to hundreds  of such sources during its mission lifetime \cite{Fan:2020zhy, Zhu:2024qpp}. Consequently, we can expect to attain an EMRI catalogue to probe their population models. The EMRI formation theories include the loss cone formation channel \cite{Amaro-Seoane:2007osp, Babak:2017tow}, the accretion disc driven formation channel \cite{Pan:2021oob}, and the supernova driven formation channel \cite{Bortolas:2019sif}.  As a first preliminary assessment, this work focuses on the widely studied loss cone formation channel and explores the constraints of TianQin imposes on their population distributions.  A more informative measure of the science ability to discriminate among these alternative population models will be addressed in future work.

The structure of EMRI population models is dictated by the physical processes and evolutionary environments in which EMRIs are expected to form and merge \cite{Babak:2017tow}, which is not sufficient to allow for a high-fidelity validation at present.  
As a first step, we ignore the detailed formation history used in EMRI population synthesis. Instead, we introduce a simple parametrization method designed to capture the salient features of the population models. Here, we are interested in investigating how effectively TianQin will resolve the distribution shape of the EMRI population models with these salient features and how accurately the population distribution shape can be recovered with the TianQin detectable EMRI events. 
 
To reconstruct the population distributions from the incomplete observed EMRI sources \cite{Mandel:2018mve, Gair:2022fsj, Thrane:2018qnx} and infer the attendant astrophysical model responsible for them \cite{Pan:2021lyw, Babak:2017tow}, a hierarchical Bayesian method is generally employed \cite{Ruan:2018tsw, Adams:2012qw}. 
This method handle the analysis on two levels: one to consider the space of the population models, and another to consider the parameters of the models themselves. Its likelihood is a joint distribution that describes the probability of obtaining the EMRI detectable catalog, given the hyper-parameters that describe the population model and the source parameters under this model. Due to the GW detector noise, the EMRI detectable catalog only contains sources loud enough to surpass the detection threshold. This introduces model selection bias \cite{Adams:2012qw, Farr:2019rap,Tiwari:2017ndi} and should be considered into populaiton analysis in order to accurately determine the true population distributions \cite{Mandel:2018mve}.

This paper attempts to obtain the posterior of the hyper-parameters with the EMRI detectable catalog. 
  Using the population model given by \cite{Babak:2017tow} as input, and applying the analytic kludge (AK) method \cite{Barack:2003fp} to map the EMRI parameters to the waveforms, the EMRI detectable catalog can be obtained with signal-to-noise ratio calculations \cite{Fan:2020zhy}. To address the selection effects, one could optimize the hierarchical Bayesian model by determining the fraction of the proposed population that is detectable and re-weighting the population likelihood accordingly. Due to the existence of noise, we can not have a perfect measurement of the parameters for a given event. The standard Bayesian method \cite{Ye:2023lok, Babak:2009ua, Chua:2021aah} to give the probability, that observing the EMRI event with specific source parameters, is too costly for population studies. For simplicity, we employ the Fisher information matrix \cite{Finn:1992wt, Vallisneri:2007ev} to estimate the parameter precision.  Then, the probability density distribution of the EMRI parameters for a given event can be approximated by a multivariate normal distribution, with the true values as the means and the Fisher results as the variances.

This paper is organized as follows: In the Sec.\ref{method}, we describe the method of hierarchical Bayesian inference. In the Sec.\ref{numberSet}, we describe the numerical setup, which include the population models, the TianQin detectable EMRI events, the selection bias and the Fisher information matrix. In the Sec.\ref{result} and Sec.\ref{conclusion}, we present our result and conclusions. 

\section{Method}\label{method}

TianQin is expected to detect tens to hundreds of EMRI sources in the future \cite{Fan:2020zhy}. With this large number of sources, it will become possible to study the populations properties.  A feasible method for this is the hierarchical Bayesian inference \cite{Ruan:2018tsw}, which allows us to go beyond individual events to study broader population properties.

The EMRI population properties can be described by a set of hyper-parameters $\vec\lambda$. Assuming the observed EMRI catalog by TianQin constitutes the data set $\{\vec{d}_i\}$, the posterior probability of $\vec\lambda$ will be given by the Bayesian formalism \cite{Mandel:2018mve}
\begin{equation}\label{equ:p_post}
p(\vec{\lambda}|\{d_i\})=\frac{p(\{d_i\}|\vec{\lambda})\pi(\vec{\lambda})}{p(\{d_i\})},
\end{equation}
where ${d_i}$ is the $i$-th event in the EMRI detectable catalog, $\pi(\vec\lambda)$ is the hyper-prior, 
$p(\{d_i\}|\vec\lambda)$ is the likelihood of observing the detectable catalog given the population properties,  and $p(\{d_i\})$ is the evidence, which can be regarded as a normalization constant and does not need an explicit calculation in the data analysis.

The $i$-th event $d_i$ is related to its source parameters ${\vec{\theta}}$ with the likelihood $p({d}_i|\vec{\theta})$, and the source parameter distribution under the population model with hyper-parameters $\vec{\lambda}$ is $p(\vec{\theta}|\vec{\lambda})$, then the likelihood $p(\{d_i\}|\vec{\lambda})$ is described as
\begin{equation}\label{equ:lam_like}
\begin{aligned}
p(\{d_i\}|\vec{\lambda})&=\prod\limits_{i=1}^{N_{\rm obs}}\frac{\int {\rm d} \vec{\theta} p(d_i|\vec\theta)p(\vec{\theta}|\vec{\lambda})}{\int_{d_i>\rm threshold} {\rm d} {d_i} \int {\rm d} \vec{\theta} p(d_i|\vec\theta)p(\vec{\theta}|\vec{\lambda})},\\
&\times e^{-N_{\rm det}}(N_{\rm det})^{N_{\rm obs}}\\
&=\prod\limits_{i=1}^{N_{\rm obs}}\frac{\int {\rm d} \vec{\theta} p(d_i|\vec\theta)p(\vec{\theta}|\vec{\lambda})}{\int {\rm d} \vec\theta p_{\rm det}(\vec\theta)p(\vec{\theta}|\vec{\lambda})}\times e^{-N_{\rm det}}(N_{\rm det})^{N_{\rm obs}}\\,
\end{aligned}
\end{equation}
where $N_{\rm obs}$ is the event number in the EMRI detectable catalog, $\int {\rm d} \vec\theta p_{\rm det}(\vec\theta)p(\vec{\theta}|\vec{\lambda})$ is the normalization factor  accounting for the overall probability given a particular choice of $\vec\lambda$. Here, $p_{\rm det}(\vec\theta)$  is the detection probability for parameters $\vec\theta$, which incorporates the selection bias that some events are easier to be detected than others, $N_{\rm det}$ is the expected number of detections once selection effects are included.

Based on Eqs. (\ref{equ:p_post}) and (\ref{equ:lam_like}), we can recover the posterior probability distribution of $\vec\lambda$ with the Markov Chain Monte Carlo (MCMC) techniques.

\section{numerical setup}\label{numberSet}
\subsection{Population Models}\label{Sec:popModel}

 Ignoring the spin of the CO, an EMRI is generally characterized by its seven intrinsic parameters: the MBH mass, the CO mass, the MBH spin, the orbital eccentricity, the orbital inclination angle, and the two phase angles describing the pericenter precession and the Lense-Thirring precession, and its seven extrinsic parameters: redshift, plunge time, two spin orientation angles, two sky oriention angles and the initial orbital phase. We refer to the set of these parameters as $\vec\theta$. The population model, $p(\vec{\theta}|\vec\lambda)$, describes the distribution of EMRI parameters. 
 %are determined by the astrophysical processes that form EMRIs.
Nonparametric models to fit the complex structures of EMRI population distributions, which are inherent in astrophysical formation processes, will likely face computationally costly and larger uncertainties in future analyses. A more practical approach, due to its simplicity, is to parametrize the rate density using basic functions and to infer their parameters. As an initial attempt, we try to fit a parameterized model that captures the simple
large-scale features of EMRI distributions. More deep exploration could be considered for future work.

%Additionally, the parameterized model could also serve as a guide to the real nonparametric models to be deep explored to interprete the theoretical EMRI formation channels.
Composed of MBHs and COs, EMRI population can be sampled from the product of MBH mass function and the accretion rate of MBHs with respect to COs. However, EMRI formation should satisfy certain necessary conditions. For example, MBHs should be located in galaxies where they are surrounded by a cusp of stars and COs, which thus serve as nurseries for EMRI formation. Moreover, correction factors should be considered to ensure that the MBHs do not overgrow their present masses by capturing too many EMRIs and plunges. The hyper-parameters $\vec\lambda$ accounting for these sophisticated realistic corrections have not yet been accurately obtained. Here, we explore the parametrization method to capture the features of the EMRI event catalog, and provide the numerical values $\vec\lambda$ to describe these features. To test the correlation between population parameters in the simulated EMRI event catalog, we computed the Spearman rank correlation coefficient $\rho$ \cite{Heinzel:2024hva}. The results indicate that $\rho=-0.013$ between redshifts and MBH masses and remain small for other parameters, suggesting that a non-correlation assumption could be reasonable. Basing on these finding, we explored the MBH mass distribution, the redshift distribion and the spin distribution, details are listed as below

\begin{itemize}
\item MBH Mass Distribution

In both the semi-analytic model and the empirical model, the MBH mass function is represented as ${\rm d}N/{\rm d}\log M\propto M^{\alpha}$. This corresponds to a number density function of ${\rm d}N/{\rm d}M\propto M^{\alpha-1}$. If we assume that each MBH with mass $M$ has an equal probability of being an EMRI,  the probability density distribution $p(M)$ in the EMRI population model would be expected to follow a one-parameter power-law.  However, due to the correction factors mentioned earlier, the distribution of $p(M)$ has been altered. In this paper, we adopt the MBH mass function follows \textit{Model pop III} \cite{Babak:2017tow}, which features a negative index of $\alpha=-0.3$. Consequently, the number of MBHs decreases with increasing MBH mass. To prevent MBHs from excessive growth, MBHs with smaller mass are assigned with larger correction factors
to reduce their EMRI formation rate. This adjustment makes $p(M)$ resembles a broken power-law, which is characterized by the following formalism

\begin{widetext}
\begin{equation}
p(M|\alpha_1,\alpha_2, b, M_{\rm min}, M_{\rm max} )\propto\left\{
\begin{aligned}
&{\mathcal N}_1M^{\alpha_1-1}~~M_{\rm min}\leq M\leq M_{\rm break}, \\
&{\mathcal N}_2M^{\alpha_2-1}~~{M_{\rm break}<M\leq M_{\rm max}},\\
&0 ~~~~~~~~~~~~~~{\rm otherwise}.
\end{aligned}
\right.
\end{equation}
And
\begin{equation}
\log M_{\rm break}=\log M_{\rm min}+b(\log M_{\rm max}-\log M_{\rm min}),
\end{equation}
\end{widetext}
where $M_{\rm min}$ and $M_{\rm max}$ represent the minimum and maximum MBH masses that are within the TianQin detectable mass range, respectively. $M_{\rm break}$ is the mass at which there is a break in the distribution spectral index, and $b$ is the fraction of the way between $M_{\rm min}$ and $M_{\rm max}$ at which the MBH distribution undergoes a break. ${\mathcal N}_1$ and ${\mathcal N}_2$ are parameters for normalizing the probability density distribution. In this paper, we focus on the constraints of TianQin on the hyper-parameters $\alpha_1, \alpha_2, b$. We expect these recovered parameters will provide the typical characteristic of this loss cone model.

In this model, $M_{\min}=3\times10^3M_\odot$ and $M_{\rm max}=10^7M_\odot$, which encompass the most sensitive MBH sources for the TianQin detector. By fitting to the EMRI event catalog of M1 in \cite{Babak:2017tow}, the hyper-parameters $\alpha_1,\ \alpha_2,\ b$ are determined to be $0.7,\ -0.98,\ 0.5$, respectively.

\item  Redshift Distribution

$p(z)$ represents the average number density of EMRIs per time as a function of redshift.  One could assume that the EMRI population remains constant with comoving volume, which implies that galaxies contribute a constant EMRI formation rate over cosmic history. However, the correction factors like the cusp regrowth time, which affects the galaxy being a nursery
for EMRI formation, will influence the redshift distribution. A simple description, that only include the form of the redshift distribution $p(z)$ in the event catalog, exhibites a truncated normal distribution. Since the normal distribution does not adequately describe the redshift distribution at high redshift, we explore another mixed power-law and Gaussian model. We present the first model with formalism as,
%\item[$\blacktriangleright$]
\begin{equation}\label{equ:redmodel1}
p(z|\mu_z,\sigma_z)=G(\mu_z,\sigma_z) ~~(z_{\rm min}<=z<=z_{\rm max}),
\end{equation}
where $z_{\rm min}$ and $z_{\rm max}$ are the minimum and the maximum redshift in the event catalog, $\mu_z$, $\sigma_z$ are the mean and the width of the $z$ distribution, respectively. And we present the second model with formalism as, 
%\item[$\blacktriangleright$]
\begin{equation}\label{equ:redmodel2}
\begin{aligned}
p(z|\lambda, \alpha, \mu_z,\sigma_z)=&(1-\lambda)\cdot \mathcal{A}_1 \cdot(1+z)^\alpha\\
&+\lambda\cdot \mathcal{A}_2\cdot G(\mu_z,\sigma_z),
\end{aligned}
\end{equation}
where $\lambda$ is the mixing fraction of the relative prevalence of the first term and the second term. $\alpha$ is the spectral index. $\mathcal{A}_1$ and $\mathcal{A}_2$ are the normalizing parameters.

The EMRI events catalog is truncated beyond redshift $z = 4.5$ to increase the calculation efficiency, otherwise, the simulated results will be dominated by undetectable sources \cite{Fan:2020zhy}.
We then fit the simulated catalog and determine the hyper-parameters. For the first redshift model, $\mu_z$ and $\sigma_z$ corresponds to 2.69 and 1.35, respectively.  For the second redshift model, $\lambda$, $\alpha$, $\mu_z$ and $\sigma_z$ corresponds to 0.78, 1.9, 2.38 and 1.12, respectively.
%The truncating redshift is determined by a conservative value of EMRI horizon distance with TianQin.

\item Spin Distribution

GW observations of EMRIs will disclose information about how the MBHs are spinning and provide insights into how and where the MBHs form. In the loss cone model, the spins of MBHs have near-maximal values. This is because MBH seeds need to accrete a sufficient amount of mass in order to enter the sensitivity band of space-borne detectors. During this process, they accumulate spin from their accreted material. 

In practice, we adopt a flat distribution over a small range of high spins
\begin{equation}
p(a)={\rm contant}, ~ (a_{\rm min}<a<a_{\rm max})
\end{equation}
where $a_{\rm min}, a_{\rm max}$ are the minimum and the maximum spin values, 
 corresponding to 0.96 and 0.998 \cite{Thorne:1974ve}, respectively.
 
For spin distribution of EMRIs, there are no hyper-parameters.

\end{itemize}
 
Besides the MBH mass, the redshift, and the MBH spin, the CO masses are assumed to be $10M_\odot$. For the other parameters, the inclination, the sky position, and the spin orientation angles are assumed to be distributed isotropically on the sphere. The three phase angles, which are uninformative for the EMRI waveforms, are assumed to be uniformly distributed between [0, $2\pi$]. Plunge time is taken to be uniform in [0, 5] yr, and eccentricities are taken to be uniform in [0, 0.2], as a rather flat distribution at the plunge is simulated in the loss cone model. These parameters, which influence the detectability of EMRI events, don't have hyper-parameters. Here, to simplify the selection bias calculation in Sec.\ref{Sec:selBias}, we assume the impact of these parameters without hyper-parameters to be a constant that reduces the number of  detectable events. This factor is defined as the ratio of the detectable event number under the above assumed conditions to the optimal conditions. We expect these selection bias can be effectively corrected using the method proposed in \cite{Chapman-Bird:2022tvu} and will not alter the result very much.

\subsection{TianQin detectable EMRI events}\label{Sec:DetNum}

The strength of a GW signal in the detector can be characterized by the signal-to-noise ratio (SNR). TianQin detectable EMRI catalog collects those EMRI events that have an SNR greater than the detection threshold. The number of EMRI events in the catalog is denoted by $N_{\rm obs}$.
Define the noise-weighted inner product between two signals $x(t)$ and $y(t)$ as
\begin{equation}
\langle x|y \rangle=4\Re \int^{\infty}_0 \frac{\tilde{x}^*(f)\tilde{y}(f)}{S_n(f)}{\rm d}f,
\end{equation}
where $\tilde{x}(f), \tilde{y}(f)$ are the Fourier transforms of $x(t)$ and $y(t)$, $S_n(f)$ is the one-sided power spectral density of the TianQin detector noise with
\begin{equation}
\begin{aligned}
S_n(f)=&\frac{1}{L^2}\Big[\frac{4S_a}{(2\pi f)^4}\Big(1+\frac{10^{-4}\rm Hz}{f}\Big)+S_x\Big]\\
&\times\Big[1+0.6\Big(\frac{f}{f_*}\Big)^2\Big],
\end{aligned}
\end{equation}
where $L=1.7\times10^8$m is the armlength of the TianQin detector, $S^{1/2}_a=1\times10^{-15}\rm m\cdot s^{-2}/Hz^{1/2}$ and $S^{1/2}_x=1\times10^{-12}\rm m/Hz^{1/2}$ are the residual acceleration noise and position noise, respectively, $f_*=1/(2\pi L)$ is the transfer frequency. Then, the optimal SNR accumulated in the observation time can be defined as:
\begin{equation}\label{equ:snr}
\rho_{\rm opt}=\langle h|h \rangle^{1/2},
\end{equation}
where $h(t)$ is the EMRI signal with the detector response.

 \begin{figure*}
\centering
\begin{minipage}{0.45\linewidth}
\includegraphics[width=0.9\columnwidth,clip=true,angle=0,scale=1.1]{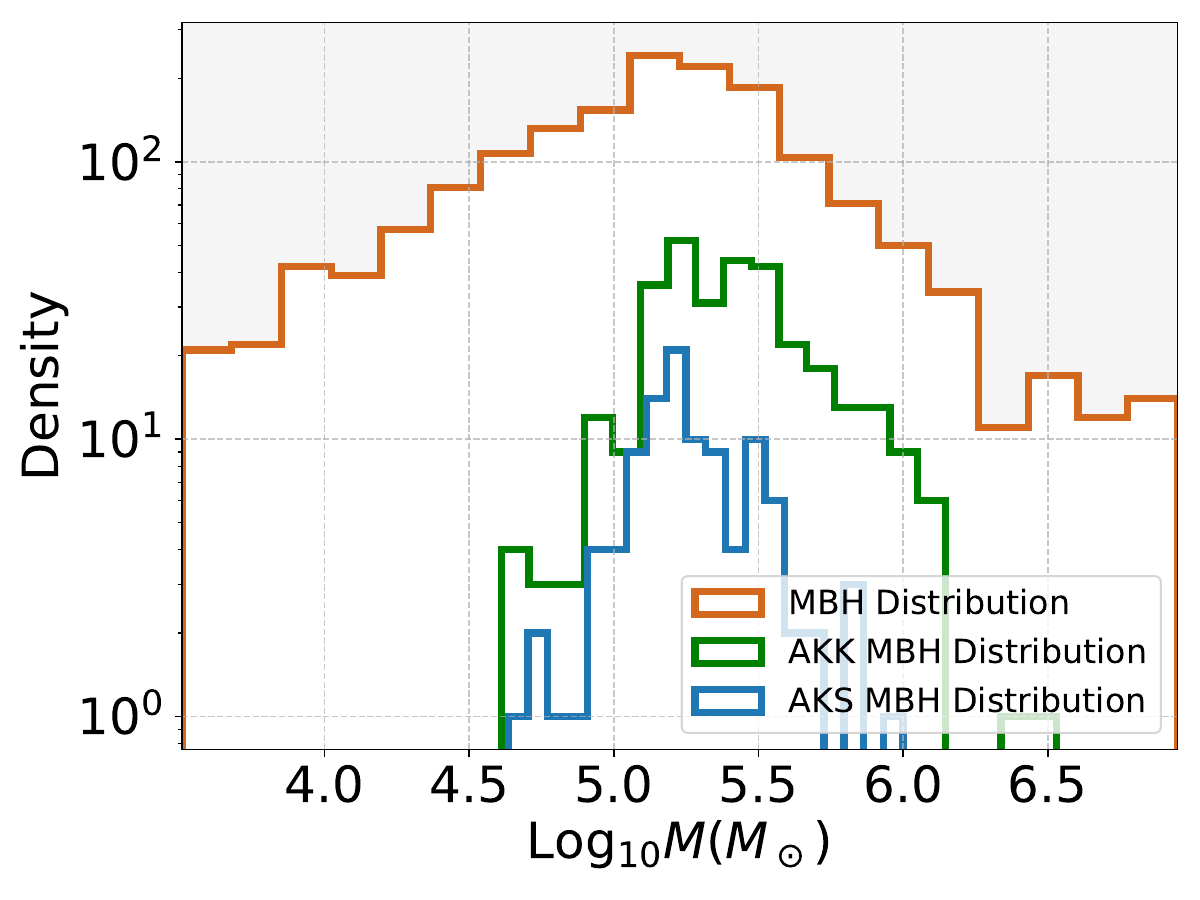}
\end{minipage}
\begin{minipage}{0.45\linewidth}
\includegraphics[width=0.9\columnwidth,clip=true,angle=0,scale=1.1]{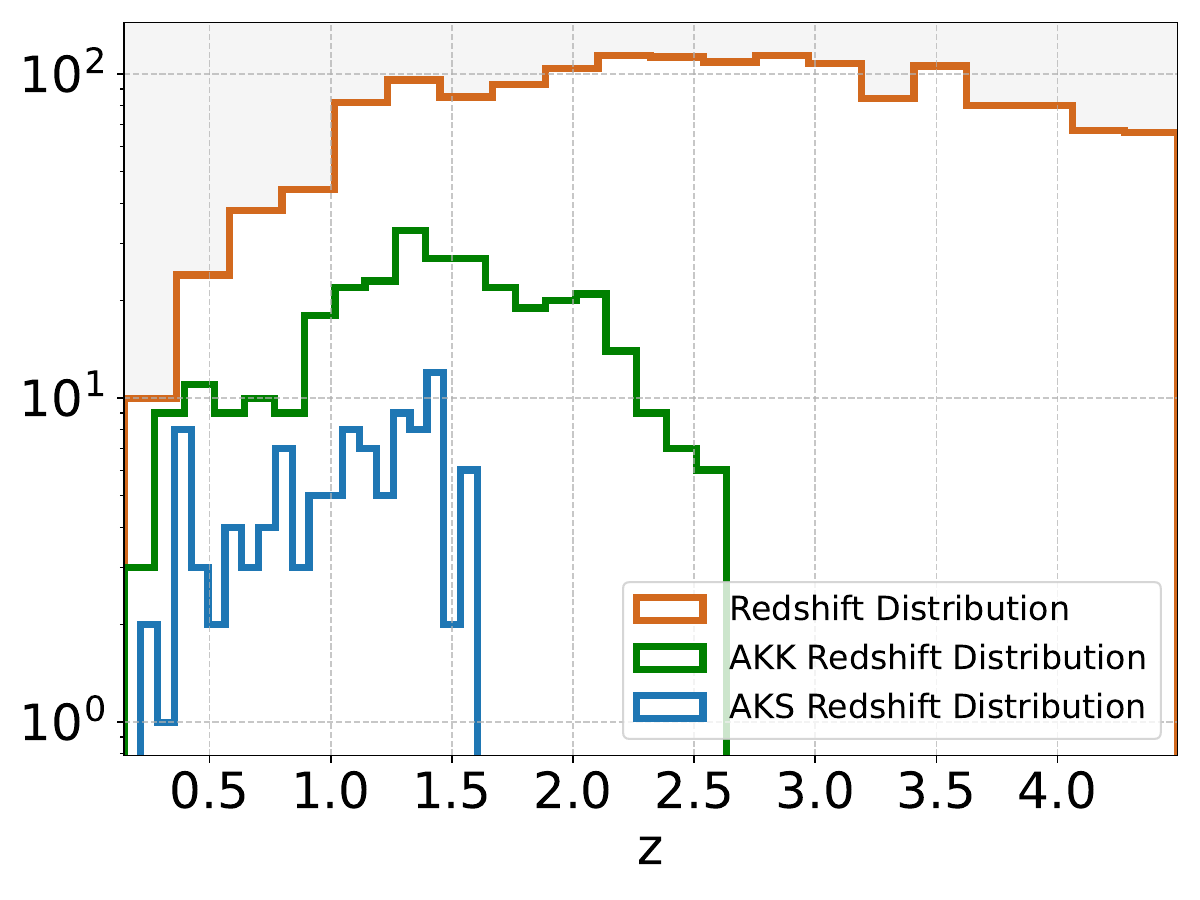}
\end{minipage}
\caption{The MBH mass and the redshift distribution in the event catalog (yellow) and their detectable distribution under the AKS (blue) and AKK (green).}
\label{fig:DetMZDis}
\end{figure*}

Using the population model described in Sec.\ref{Sec:popModel}, and the EMRI event catalog according to the underlying distributions, the EMRI waveforms can be calculated. Here, we employ the AK method  \cite{Barack:2003fp} to access the EMRI's waveform. This is for us to facilitate comparison with the previous work \cite{Fan:2020zhy}, although several more efficient waveform methods are proposed. In the AK method \cite{Barack:2003fp}, the two polarizations of the EMRI waveform are defined as
%\begin{widetext}
\begin{equation}
\begin{split}
h^+=&\sum_n-\big[1+(\hat{L}\cdot\hat{n})^2\big]\big[a_n \cos2\gamma-b_n \sin2\gamma\big]\\
&+\big[1-(\hat{L}\cdot\hat{n})^2\big]c_n, \\
h^\times=&\sum_n 2(\hat{L}\cdot\hat{n})\big[b_n\cos2\gamma+a_n\sin2\gamma\big],
\end{split}
\end{equation}
%\end{widetext}
where $\hat{L}$ is a unit vector along the CO's orbital angular momentum, $\hat{n}$ is the unit vector pointing from the detector to the source, $\gamma$ is the azimuthal angle measuring the direction of the pericenter, $a_n, b_n, c_n$ are combination of three independent components, which are the second time derivative of the inertia tensors calculated based on the Fourier analysis of the Kepler orbit. More details can be found in \cite{Barack:2003fp}. The EMRI signals entering the TianQin detector cause a shift of  the armlength, and generate response signals with strain amplitude as follows
\begin{equation}
h(t)=\frac{\sqrt{3}}{2}\big[F^+(t)h^+(t)+F_\times(t)h^\times(t)\big],
\end{equation}
where $F^+(t), F^\times(t)$ are the antenna pattern, whose detailed expressions can be found in \cite{Fan:2022wio}. By substituting $h(t)$ into equation (\ref{equ:snr}), the SNR for all EMRI events can be obtained. We choose the EMRI waveforms truncated at the last stable orbit of the Schwarzschild black hole (AKS) and the Kerr black hole (AKK), and adopt the detection thresholds of 15 as suggested in \cite{Babak:2009ua} and \cite{Babak:2017tow}, respectively. Then, we count the EMRI events with SNR exceeding these thresholds. We find that about 100 EMRI events can be detected for AKS waveform, and about 320 EMRI events could be detected for AKK waveform, during TianQin mission lifetime.

%as the AK method for EMRI waveform is more reliable when the compact object (CO) is far away from the plunge

We present a comparison of the original population and the observed population in Fig.\ref{fig:DetMZDis}, with a detection threshold of 15. The left plot of Fig.\ref{fig:DetMZDis} illustrates the distribution of the MBH mass, while the right plot of Fig.\ref{fig:DetMZDis} shows the distribution of the redshift. These distributions result from the original source distributions, the frequency-dependent sensitivity curve, and the correlation between MBH mass and the peak frequency of its gravitational wave (GW) signal. 
In the left plot of Fig.\ref{fig:DetMZDis}, the results indicate that the mostly observed MBH masses ( the green and blue color) are distributed between  $3\times10^4M_\odot$ and $2\times10^6M_\odot$, with the most detectable MBH mass being around $2\times10^5M_\odot$. 
 EMRI sources with an MBH mass beyond this range, which cannot be observed, will introduce a significant selection bias. In the right plot of Fig.\ref{fig:DetMZDis}, the observed redshifts (the  green and blue color) are mainly distributed at $z\leq2.6$, due to the finite detectable range of TianQin for EMRIs.  The results justifies our choice of redshift truncate at 4.5.

 %This implies that we can only expect to constrain redshift evolution within this range. Also,

 \subsection{Selection Bias}\label{Sec:selBias}
 
To extract the distribution properties of EMRI sources using the hierarchical Bayesian method, one often needs to deal with selection effects.  First, the loudest or brightest sources are more likely to be detected.  Second, there are uncertainties in the parameter measurements of individual observations. Therefore, it is necessary to correct these biases in order to determine the original source population distribution accurately.

For these selection biases, the crucial step is to include the detection probability in the normalization factor. This adjustment takes into account the different event numbers expected to be observed under varying population models. It is represented by the expression $\alpha(\vec\lambda)=\int d\vec\theta p_{\rm det}(\vec\theta)p(\vec\theta|\vec\lambda)$, as shown in equation (\ref{equ:lam_like}). To evaluate this expression, one can approximate it by performing a Monte Carlo sum with 
 \begin{equation}\label{equ:alpha}
 \alpha(\vec\lambda)=\frac{1}{N_t}\sum^{N_t}_{k=0}p_{\rm det}(\vec\theta),
 \end{equation}
 where ${\vec\theta}$ are sampled from $p(\vec\theta|\vec\lambda)$, $N_t$ is the number of samples, and $p_{\rm det}(\vec\theta)$ is the detection probability for parameters $\vec\theta$. Due to fluctuations in the detector noise, the SNR of the observed source with parameters $\vec\theta$ is not fixed. $p_{\rm det}(\vec\theta)$ is calculated based on a cut on the SNR that exceeds the detection threshold and thereby the corresponding probability from the SNR likelihood distribution.  There are different ways to calculate $p_{\rm det}$. One practical method is to express the distribution of SNR $\rho$ as a normal distribution with mean $\rho_{\rm opt}$ and unit variance. Thus, 
 \begin{equation}
 p_{\rm det}(\vec\theta)=\frac{1}{2}\rm erfc\left(\frac{\rho_{\rm th}-\rho_{\rm opt}( \vec{\theta})}{\sqrt{2}}\right),
 \end{equation}
where $\rho_{\rm th}$ is the EMRI detection threshold. In reality, to achieve a percent-level accuracy of $\alpha(\lambda)$, the sample size $N_t$ needs to be as high as $10^5$, which is infeasible for addressing the SNRs in a sampling run. To solve this problem, we calculate the horizon distance,  the farthest distance at which an EMRI source can be detected, and count the number of samples under this curve. A more rapid and accurate method can be found in \cite{Chapman-Bird:2022tvu}, which constructed an efficient neural network interpolator for selection effects calculation.

 \subsection{Fisher Information Matrix}

For the  EMRI detectable catalog, we don't have perfect measurements of the parameters of a given EMRI event. The most reliable approach to estimate the likelihood $p(d_i|\vec\theta)$ is to use Bayesian techniques with MCMC \cite{foreman2013emcee}. However, these methods are too computationally expensive to be used as the approaches to make forecasts for future observations. Instead, we approximated the Fisher information matrix (FIM) to employ the EMRI likelihood. It is a common tool to quantify the parameter measurement uncertainty, where its diagonal values represent the estimation precision for an unbiased physical parameter.

The FIM matrix is defined as 
\begin{equation}
\Gamma_{ij}=\Big(\frac{\partial\tilde{h}(f)}{\partial\theta^i}\Big|\frac{\partial\tilde{h}(f)}{\partial\theta^j}\Big),
\end{equation}
where $\theta^i$, $i=1,2,...,$ are the parameters of the EMRIs. The covariance matrix, which represents only the Cramer-Rao bound, can be obtained as 
\begin{equation}
\Sigma_{ij}\equiv\langle\delta\theta_i\delta\theta_j\rangle=(\Gamma^{-1})_{ij}.
\end{equation}
The marginal uncertainty $\sigma_i$ for the $i-$th parameter can be derived as 
\begin{equation}
\sigma_i=\Sigma_{ii}^{1/2}.
\end{equation}
Then, the likelihood $p(d_i|\vec\theta)$ can be approximated by a multivariate normal distribution with follows 
\begin{equation}
p(d_i|\vec\theta)\approx\mathcal N(\vec\theta, \Sigma_{ii}).
\end{equation} 
 
\begin{figure}
\centering
\includegraphics[width=0.9\columnwidth,clip=true,angle=0,scale=1.1]{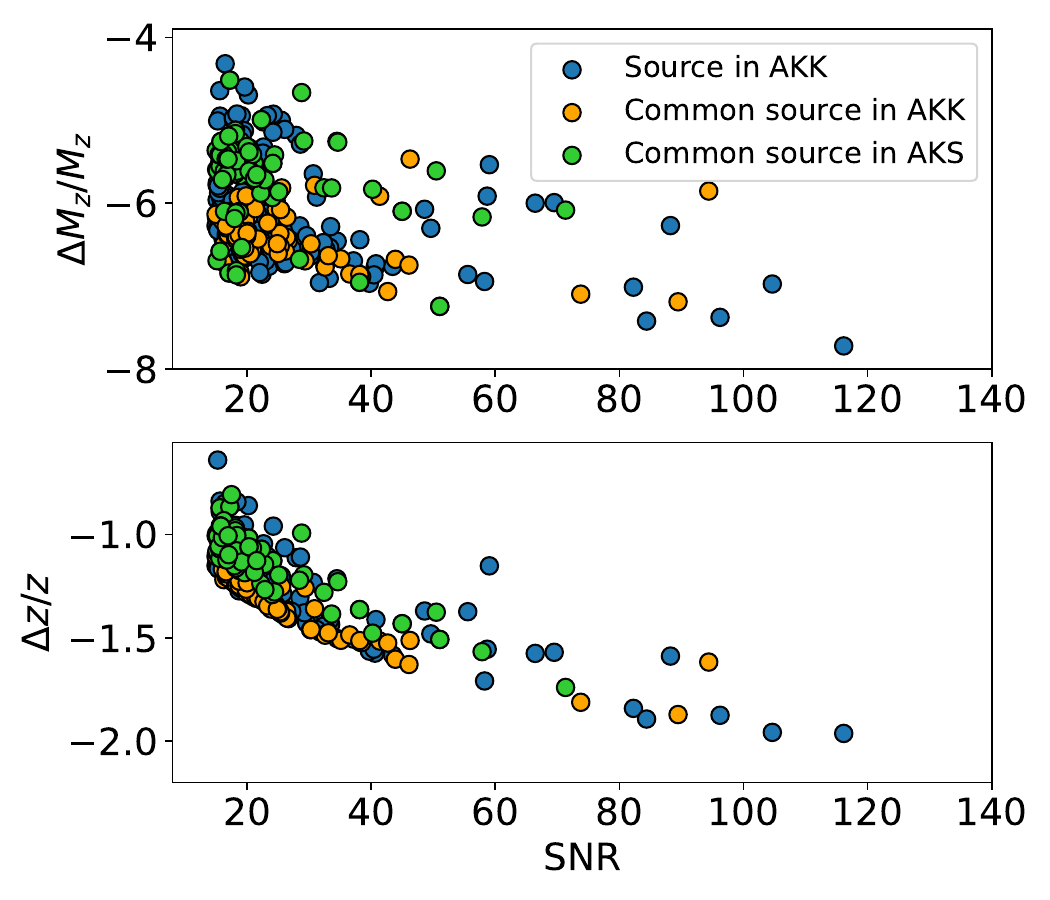}
\caption{The parameter estimation precision for the MBH mass ( upper plot) and the redshift ( lower plot) in AKS (the green points) and AKK (the blue and yellow points). }
\label{fig:ParaEsti}
\end{figure}

 The FIM-predicted uncertainties in the estimation of the EMRI parameters are given in Fig.\ref{fig:ParaEsti}. In this figure, the $y$-axis represents  the predicted error distributions, while the $x$-axis represents the SNR. The upper plot corresponds to the detector-frame MBH mass and the lower plot corresponds to the redshift. From this figure, we can find that for sources that can be detected with both AKS and AKK waveforms, the precision of parameter estimation is better with AKK waveforms, as the evolution time is longer. However, there is a large portion of events that can be detected with AKK waveforms but do not have enough SNR when using AKS waveforms. The high precision achieved with AKK waveforms is diluted by these relatively low SNR events. More detailed explanations can be found in \cite{Fan:2020zhy} and \cite{Lyu:2023ctt}. 
 
 With the results shown in Fig.\ref{fig:ParaEsti}, the 1-$\sigma$ posterior distribution for those events in the detectable catalog can be obtained.  The results demonstrate that TianQin can estimate the EMRI parameters with high precision, indicating that the likelihood $p(d_i|\vec\theta)$ is highly concentrated and closely aligned with the true parameters. The precision is calculated with the redshift mass $M_z$ in the detector-frame, which is related by $M\cdot(1+z)$ to the MBH mass sampled from the population model. 
 
 %the MBH mass in the detectable catalog should be collected in source-frame and the population inference should be under the source frame.

\section{Parameter Estimation Result}\label{result}

The EMRI population models remain quite uncertain due to a limited knowledge of their formation mechanisms. A comprehensive understanding of their distributions can be a crucial diagnostic for deriving the mechanisms that form EMRIs. In this work, we use a parametrization method to obtain the EMRI population characteristics of the loss cone model. We aim to assess to what extent the hyper-parameters can be recovered by probing the detectability of TianQin on EMRIs.

We adopt uninformative priors and specify a flat prior distribution for all the hyper-parameters.  Those describing the truncated Gaussian redshift as in equation (\ref{equ:redmodel1}) are labeled as model-1, and those describing the power-law plus Gaussian redshift as in equation (\ref{equ:redmodel2}) are labeled as model-2.  The details are listed in Table \ref{table:priors}, where the first column represents the hyper-parameters for model-1 and their true values, and the second column for model-2 and their true values, and the third column specifies their respective ranges. In this table, for model-1, the hyper-parameters that describe the slope of the MBH distribution have priors $\alpha_1, \alpha_2 \in [-10, 10]$, and  the hyper-parameter that describes the dispersion of the redshift distribution has prior $\sigma_z \in [0, 5]$, which is a  wide range relative to its true value. The hyper-parameter $b$ has a prior in the range $[0, 1]$, which is a natural condition as it describes the fraction at which the MBH distribution undergoes a break. Additionally, the prior of the hyper-parameter $\mu_z$, which describes where the average distribution number of EMRIs in redshift is centered, is assumed to be smaller than 4.5, as the population is generated below this value. For model-2, the priors for hyper-parameters $\alpha_1, \alpha_2, b, \mu_z, \sigma_z$ are same as for model-1. The $\lambda$, describing the mixing fraction of the relative prevalence of the power-law and the Gaussian terms, has a prior in the range $[0, 1]$. Additionally, the $\alpha$, describing the slope of the power-law term, has a prior in the range of $[0, 10]$.

\begin{figure}
\centering
\begin{minipage}{0.9\linewidth}
\includegraphics[width=0.9\columnwidth,clip=true,angle=0,scale=1.1]{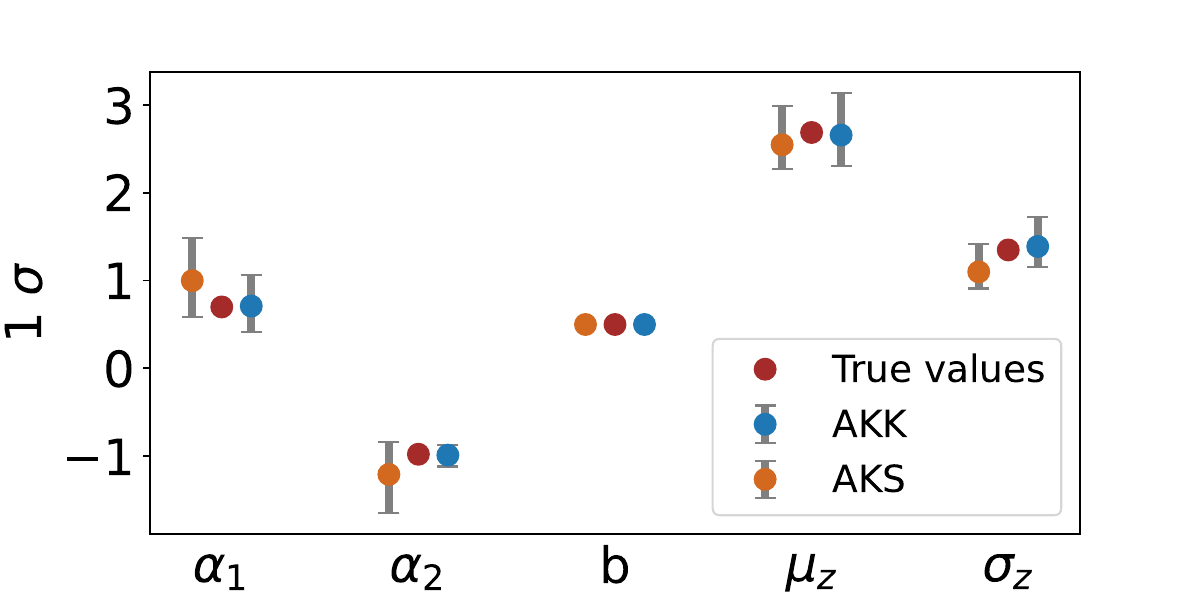}
\end{minipage}
\begin{minipage}{0.9\linewidth}
\includegraphics[width=0.9\columnwidth,clip=true,angle=0,scale=1.1]{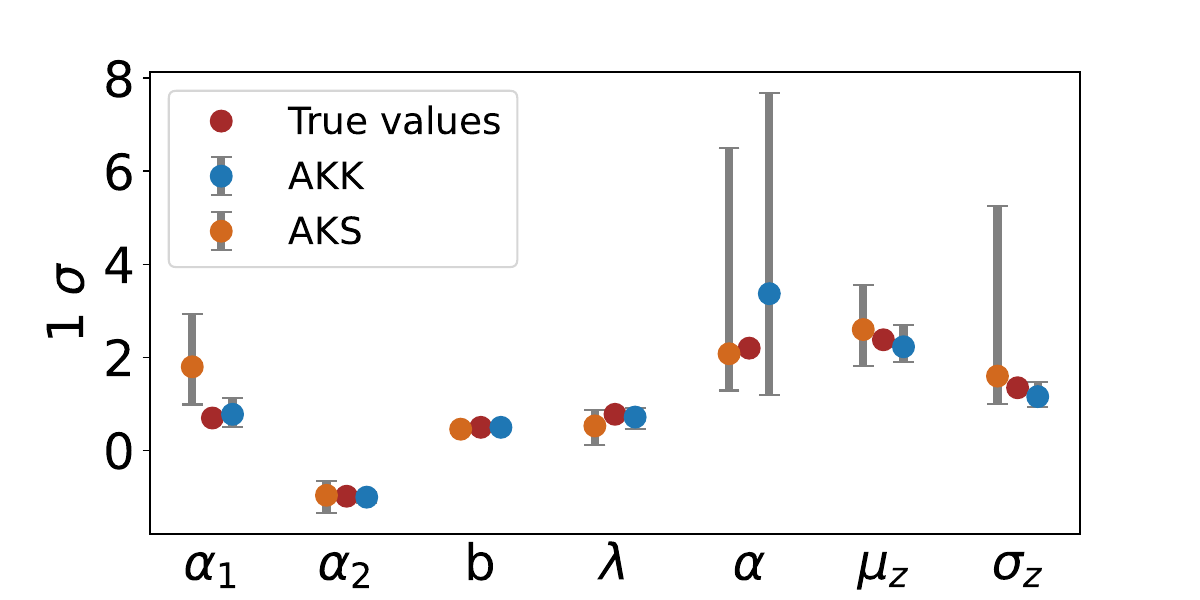}
\end{minipage}
\caption{The red dots represent the true values injected in the population model, while the yellow dot-lines and the blue dot-lines denote the 1$\sigma$ credible intervals of the hyper-parameters  with AKS and AKK waveform, respectively. The upper plot displays the result for model-1, and the lower plot shows the result for model-2.}
\label{fig:HyperParaEsti}
\end{figure}

\begin{figure*}[htbp]
\centering
\begin{minipage}{0.9\linewidth}
\includegraphics[width=0.9\textwidth,clip=true,angle=0,scale=1.1]{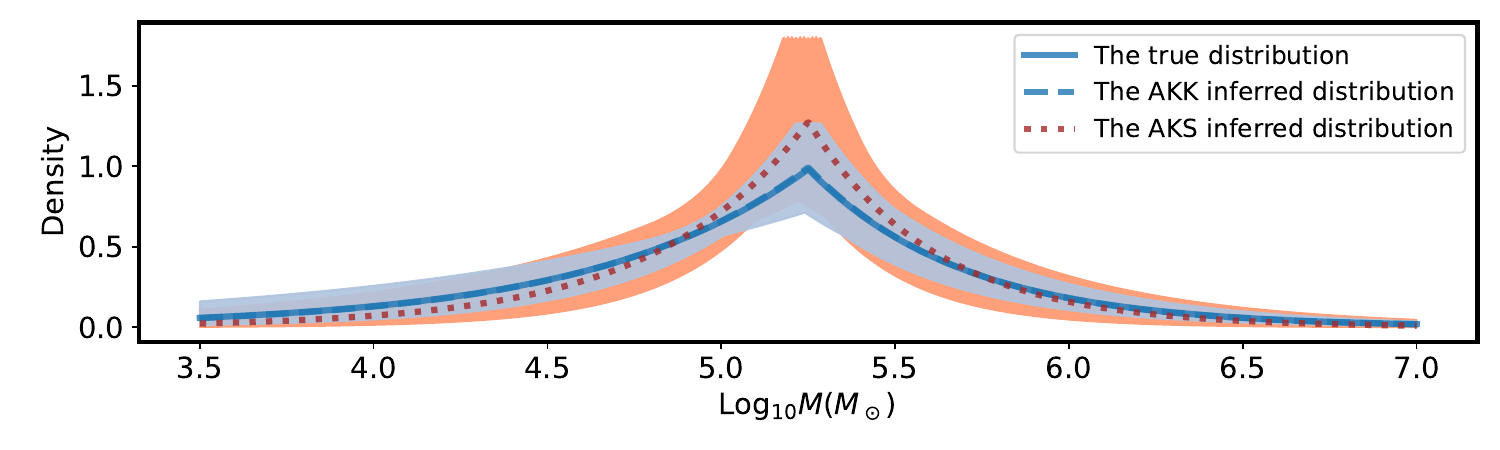}
\end{minipage}
\begin{minipage}{0.9\linewidth}
\includegraphics[width=0.9\textwidth,clip=true,angle=0,scale=1.1]{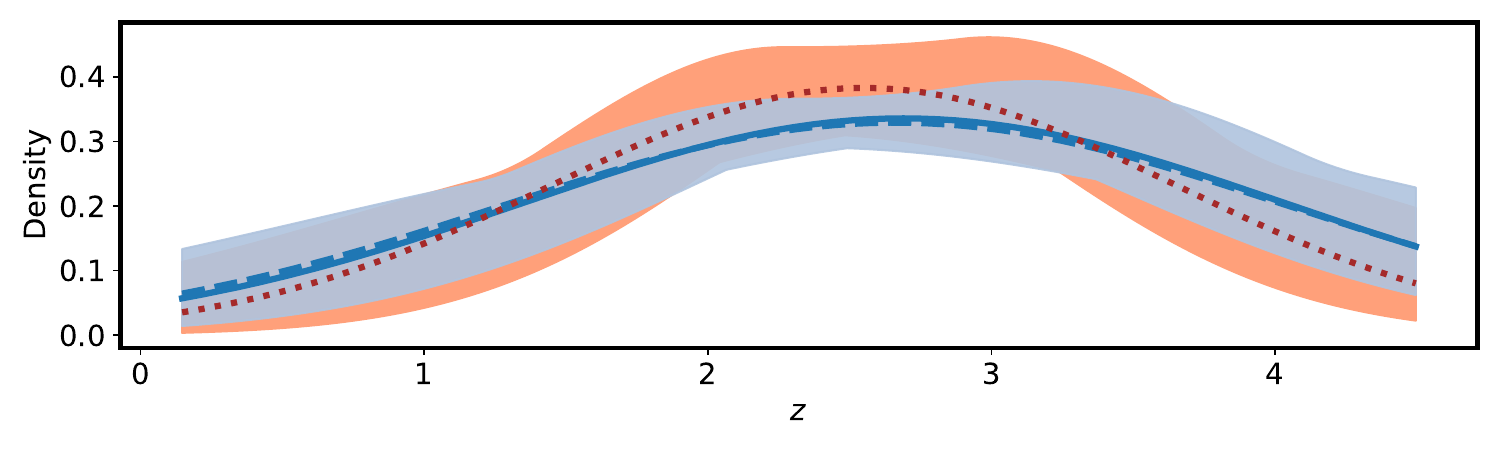}
\end{minipage}
\caption{The inferred probability density distribution of MBHs (upper plot) and redshift (lower plot) of model-1. The solid lines represent the true distributions, the dashed lines show the inferred distributions, and the shadow regions represent the $1\sigma$ credible intervals. }
\label{fig:MZdistribution1}
\end{figure*}

\begin{figure*}[htbp]
\centering
\begin{minipage}{0.9\linewidth}
\includegraphics[width=0.9\textwidth,clip=true,angle=0,scale=1.1]{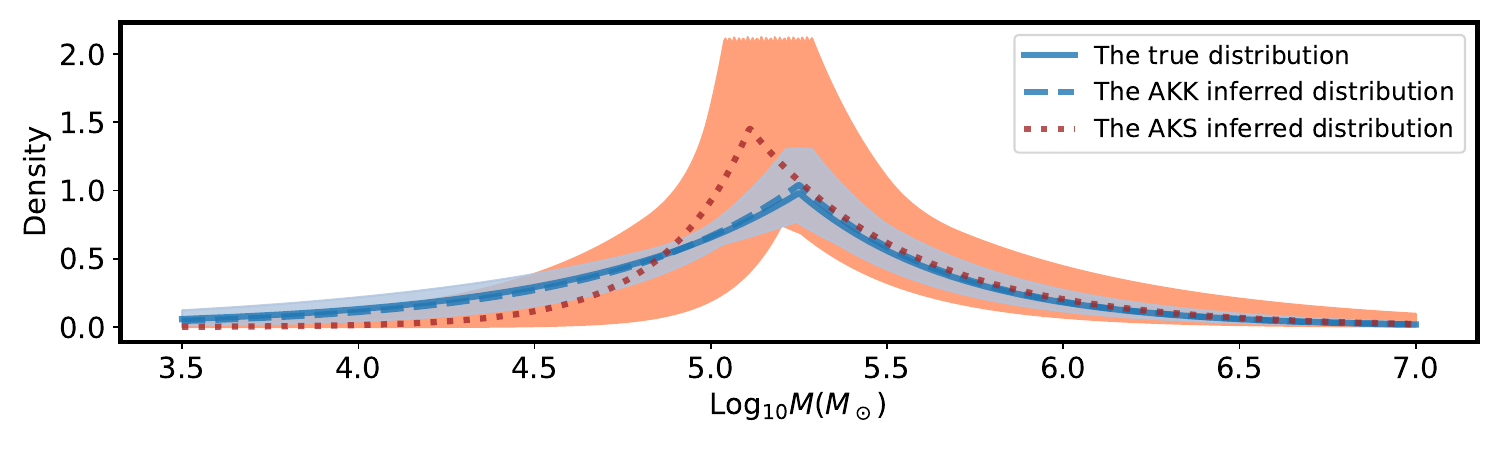}
\end{minipage}
\begin{minipage}{0.9\linewidth}
\includegraphics[width=0.9\textwidth,clip=true,angle=0,scale=1.1]{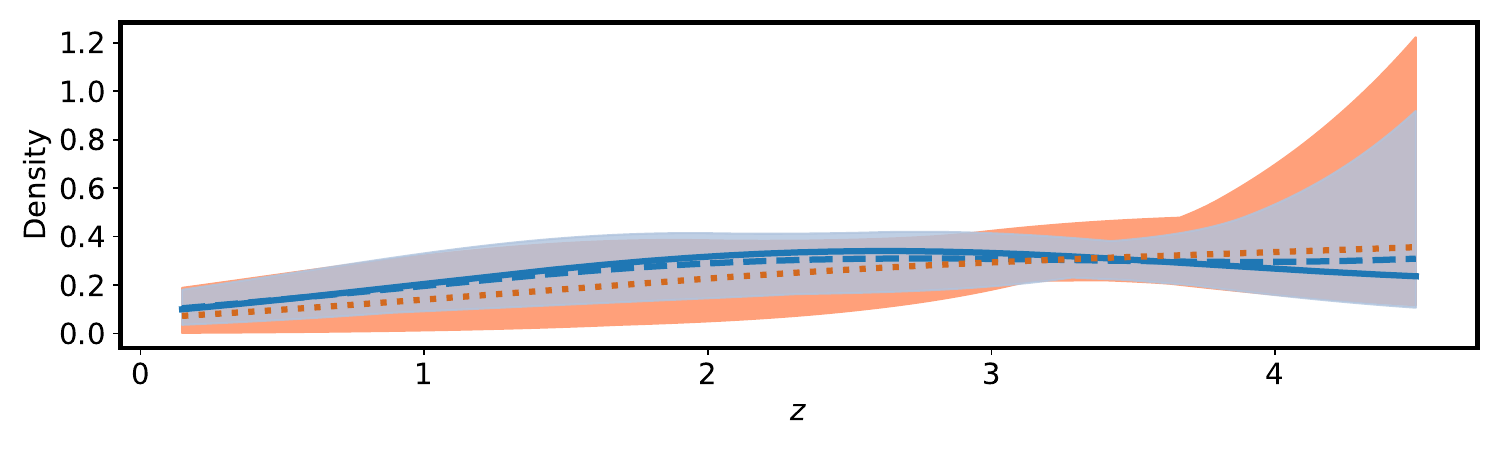}
\end{minipage}
\caption{The inferred probability density distribution of MBHs (upper plot) and redshift (lower plot) of model-2. The solid lines represent the true distributions, the dashed lines show the inferred distributions, and the shadow regions represent the $1\sigma$ credible intervals. }
\label{fig:MZdistribution2}
\end{figure*}

\begin{figure*}[htbp]
\centering
\includegraphics[width=0.9\textwidth,clip=true,angle=0,scale=1.1]{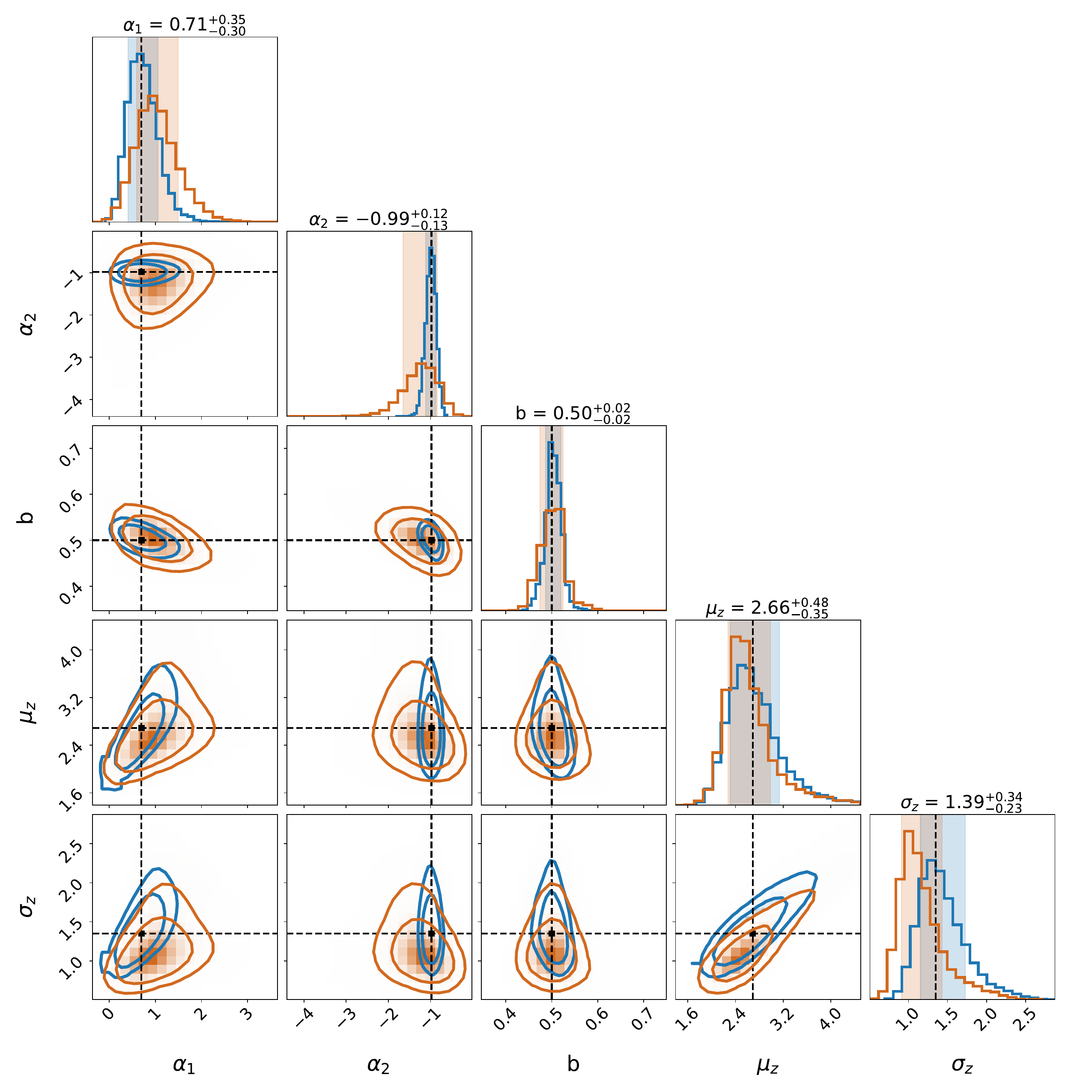}
\caption{The parameter estimation results of the hyper-parameters with AKS (yellow) and AKK (blue) detectable EMRI sources of model-1. The shadow in the posterior denotes the $1\sigma$ confidence interval.  }
\label{fig:HypePara_1}
\end{figure*}

\begin{figure*}[htbp]
\centering
\includegraphics[width=0.9\textwidth,clip=true,angle=0,scale=1.1]{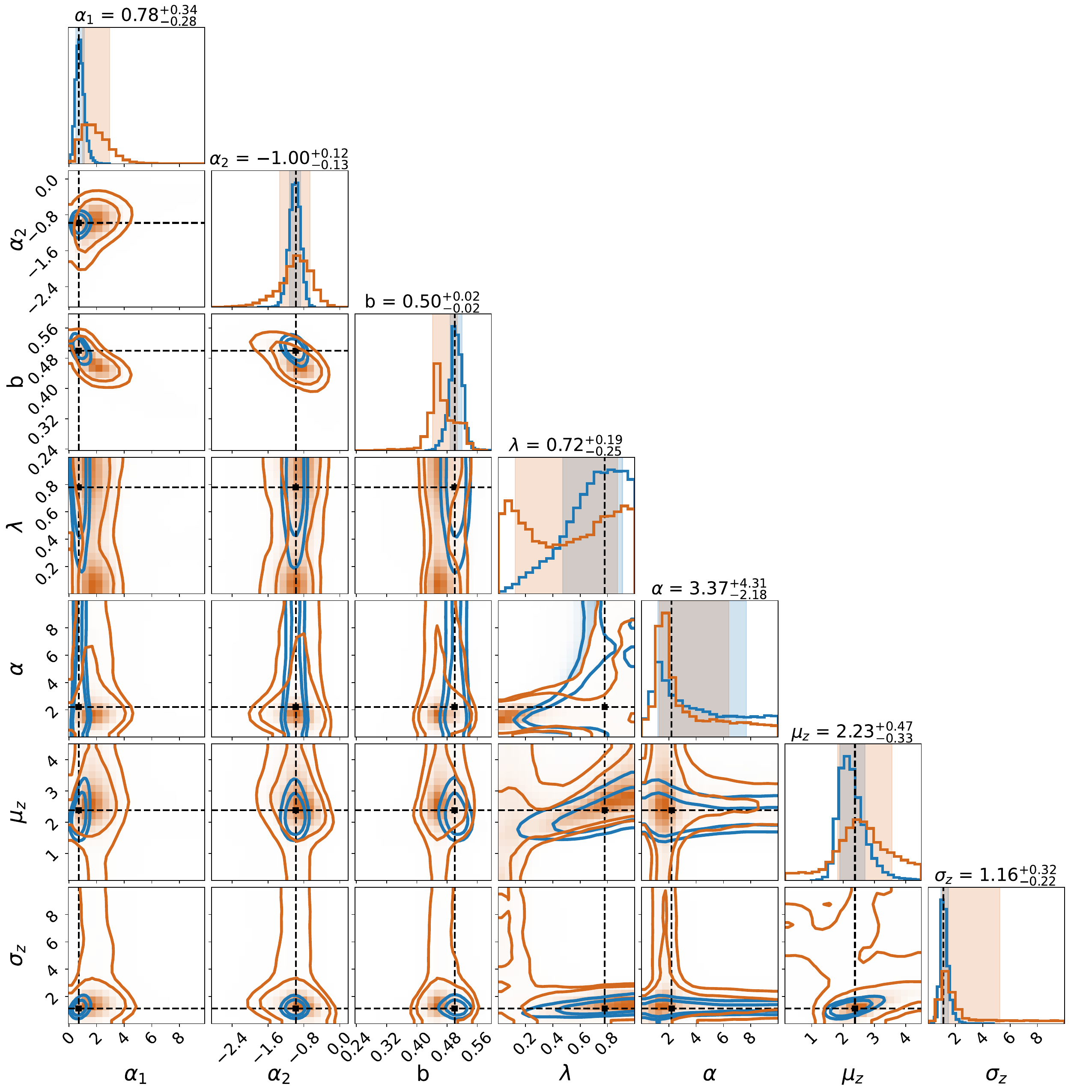}
\caption{The parameter estimation results of the hyper-parameters with AKS (yellow) and AKK (blue) detectable EMRI sources of model-2. The shadow in the posterior denotes the $1\sigma$ confidence interval.  }
\label{fig:HypePara_2}
\end{figure*}

\begin{table}[!htbp]
\caption{Priors for the EMRI hyper-parameters that describe the population models of model-1 and model-2. }
\begin{center}
\begin{tabular}{c|c|c}
\hline
Model-1 (True values)& Model-2 (True values)&Priors\\
\hline
$\alpha_1$(0.7)&$\alpha_1(0.7)$& ~~~[-10, 10]~~~\\
$\alpha_2$(-0.98)&$\alpha_2$(-0.98)&~~~[-10, 10]~~~\\
$b$(0.5)&$b$(0.5)&~~~[0, 1]~~~\\
-&$\lambda$(0.78)&~~~[0, 1]~~~\\
-&$\alpha$(2.2)&~~~[0, 10]~~~\\
$\mu_z$(2.69)&$\mu_z$(2.38)&~~~[0, 4.5]~~~\\
$\sigma_z$(1.35)&$\sigma_z$(1.12)&~~~[0, 5.0]~~~\\
\hline
\end{tabular}
\end{center}
\label{table:priors}
\end{table}
The constraint results of TianQin on these hyper-parameters are summarized in Fig.\ref{fig:HyperParaEsti} to Fig.\ref{fig:HypePara_2}. In Fig.\ref{fig:HyperParaEsti}, the red dots represent the true values, and the gray lines correspond to the error bars, which present the $1 \sigma$ credible intervals for AKS (left) and AKK (right), respectively. The upper plot denotes the result for model-1, while the lower plot denote the result for model-2. From this Figure, we can find that the hyper-parameters can be measured relatively precisely for model-1, and the posterior values are generally consistent with the true values for both the case with AKS and AKK. We can also find that the error bars of the hyper-parameters describing the MBH mass are larger in the AKS model compared to those in the AKK model. This is expected, as more EMRI sources can be detected in the AKK model. However, when comparing the error bars of the hyper-parameters describing the redshift in the upper plot of Fig.\ref{fig:HyperParaEsti}, the uncertainty range for $\mu_z$ in the AKK model is slightly larger than that of the AKS model. This arises from the correlation between the hyper-parameters $\mu_z$ and $\sigma_z$, as shown in Fig.\ref{fig:HypePara_1}, which collectively affect the parameter estimation results. In the AKK model, the median of the error bar for $\sigma_z$ lies closer to the true value, whereas in the AKS model, it deviates comparatively farther and is skewed toward smaller value. The correlation from $\sigma_z$ narrows the error bar of $\mu_z$ in the AKS model to avoid excessively low EMRI event counts at the two ends of the redshift distribution during MCMC realization, making the parameter estimation for $\mu_z$ in the AKS model appear more favorable. If the medians of the error bars for $\sigma_z$ were positioned the same, the AKK model would definitely demonstrate better results. This reminds us that ignoring the median values when interpreting error bars can lead to misleading conclusions. 

For model-2, the hyper-parameters $\lambda$ and $\alpha$ cannot be precisely recovered, especially in the AKS case. Although the bottom plot of Fig.\ref{fig:HyperParaEsti} shows a better precision result for $\alpha$ in the AKS model compared to the AKK model, we believe this is more likely to be a falsely improved result. This is because the influence of these two hyper-parameters on the redshift event distribution is minimal and more pronounced in the tail of the distribution. Since TianQin can only detect EMRI sources with redshifts smaller than 1.6 for AKS and 2.6 for AKK, respectively, the recovery of information is compromised. Consequently, the recovery of other hyper-parameters is also negatively affected. For both models, the parameter $b$ ‌has a very high estimation accuracy. Its most likely posterior value basically located at the true value and the error bar is too short to be visible in the plot of Fig.\ref{fig:HyperParaEsti}. This is because this turning point is located at the most sensitivity band of TianQin, which can also be found in the left plot of Fig.\ref{fig:DetMZDis}. From the distribution of the detectable EMRI catalog, we can find a clear break emerges at the turning point.

% is more pronounced in the tail of the redshift distribution. Since TianQin cannot detect EMRI sources with redshift larger than 1.6  and 2.6 for AKS and AKK, respectively
%while the error bars in the case with  $N_{\rm obs}=60$ are much larger compared with those in the case with $N_{\rm obs}=100$.

%This is obvious because those detected EMRI sources can be considered as samples from the population, a smaller sample size means a greater random fluctuation that will influence the parameter estimation result.

% In contrast, $\mu_z$ has a relatively low accuracy of estimation, and its most likely posterior value significantly  deviates from the true value. This is because TianQin cannot detect those EMRI sources with redshift $z>2$, which is shown in the right plot of Fig.\ref{fig:DetMZDis}. During the MCMC model search, the optimal model matching doesn't converge in redshift. In fact, EMRIs with MBH masses below $10^4 M_\odot$ and above $3\times10^6 M_\odot$ are also very difficult to be detected by TianQin. However, the hyper-parameters $\alpha_1$ and $\alpha_2$, which describe the MBH mass distribution in these regions, can be recovered very well. This is because EMRIs with MBH masses between $10^4M_\odot$ and $3\times10^6 M_\odot$ are detectable by TianQin, and their distribution tendency is the same as that in the regions mentioned before.

We present a comparison between the true and the inferred population distributions of model-1 and model-2 in Fig.\ref{fig:MZdistribution1} and Fig.\ref{fig:MZdistribution2}. In these two figures, the solid lines represent the true distributions, the dotted lines and the dashed lines indicate the inferred most likely posterior distributions for AKS and AKK, respectively, and the shadow regions represent the $1\sigma$ credible intervals. The upper plot corresponds to the MBH masses, and the lower plot to the redshift.  From Fig.\ref{fig:MZdistribution1}, we can find that the inferred MBH distribution and the redshift distribution provide a consistent result compared to the true distribution. The peak and trend of the lines are identified with high credibility within the event catalog range. This means that the characteristic of the EMRI population distributions can be well recovered, which can serve as evidence for loss cone formation channel in the future. For model-2 in Fig.\ref{fig:MZdistribution2}, 
although the inferred lines are generally consistent with the true distribution, the uncertainties are much larger compared with model-1 in Fig.\ref{fig:MZdistribution1}. 
As we explained above, this is because the first term in equation (\ref{equ:redmodel2}) has less influence on the redshift distribution of M1 in \cite{Babak:2017tow}. If more characteristics of this term are reflected in other formation channels, we believe this model will demonstrate a unique advantage in the recovery of hyper-parameters. 

%we can find a large uncertainty at the tail of the redshift distribution. 
%However, for the redshift distribution, the average distribution number, $\mu_z$, has a value that exceeds the limitation range of TianQin. This increases the errors and makes it difficult to determine this value accurately. If we adjust $\mu_z$ to match the true value, the inferred redshift distribution would be consistent with the true redshift distribution. This indicates that we could accurately determine the increasing trend of EMRI events along the redshift, while underestimating their actual number. 

For more details on the recovered hyper-parameters of model-1 and model-2, we show their posterior distributions in the corner plot of Fig.\ref{fig:HypePara_1} and Fig.\ref{fig:HypePara_2}. In these two figure,  the yellow contour and the blue contour represent the parameter estimation results of the hyper-parameters for AKS and AKK, respectively. Here, the smaller circles show the $1\sigma$ distribution range, and the larger circles represent the $90\%$ distribution range, and the black dotted lines indicate the true hyper-parameter values.  
The yellow (AKS) and blue ( AKK) shadow in the histogram denote the $1\sigma$ confidence interval, and the title for the blue shadow is shown. From the corner plot of Fig.\ref{fig:HypePara_1}, we can find that all the hyper-parameters can be recovered within $1\sigma$ confidence interval. 
For AKK, the hyper-parameters $\alpha_1, \alpha_2, b$, which describe the MBH mass distribution, has the corresponding limits $\alpha_1=0.71^{+0.35}_{-0.30}$, $\alpha_2=-0.99^{+0.12}_{-0.13}$, and $b=0.5^{+0.02}_{-0.02}$.  The hyper-parameters $\mu_z$ and $\sigma_z$, which describe the redshift distribution, has the corresponding limits $\mu_z=2.66^{+0.48}_{-0.35}$ and $\sigma_z=1.39^{+0.34}_{-0.23}$. The estimation accuracies for $\alpha_1, \alpha_2$, and $b$ are $46.4\%$, $12.6\%$ and $3\%$, respectively. The hyper-parameter $\mu_z$ could be measured with an accuracy of $15.4\%$ and the $\sigma_z$ could be measured with an accuracy of $21.1\%$. 
%which exhibits a non-Gaussian, incrementally distributed posterior as shown in Fig.\ref{fig:HypeParasnr_1}, has value above  the detectable range of TianQin
% the golden contour and the green contour represent the parameter estimation results of the hyper-parameters for AKS and AKK, respectively. The golden (AKS) and green ( AKK) shadow in the histogram denote the $1\sigma$ confidence interval, and the title for the green shadow is shown.
From the corner plot of Fig.\ref{fig:HypePara_2},  we can find that $\lambda$ cannot be completely determined in AKS. As a result, the other parameters have a very wide distribution, which greatly increases the uncertainty of hyper-parameter recovery. Compared with AKS, AKK has a much better performance in this model. For the hyper-parameters $\alpha_1, \alpha_2, b$, the corresponding limits are $\alpha_1=0.78^{+0.34}_{-0.28}$, $\alpha_2=-1.00^{+0.12}_{-0.13}$, and $b=0.5^{+0.02}_{-0.02}$. For the hyper-parameters $\lambda$, $\alpha$, $\mu_z$ and $\sigma_z$, the corresponding limits are $\lambda=0.71^{+0.19}_{-0.25}$, $\alpha=3.37^{+4.31}_{-2.18}$, $\mu_z=2.24^{+0.47}_{-0.33}$ and $\sigma_z=1.16^{+0.32}_{-0.22}$. The estimation accuracies for $\alpha_1, \alpha_2$, and $b$ are $45.0\%$, $12.7\%$ and $4\%$, respectively. In comparison, the hyper-parameter $\lambda$, $\alpha$, $\mu_z$, $\sigma_z$ could be measured with an accuracy of $28.2\%$, $147.5\%$, $16.8\%$ and $24.1\%$. With these estimation accuracy, the EMRI population characteristics can be effectively demonstrated. 

The above results show that the posterior peaks lie near the true values. However, this presentation is based on a single MCMC realization and is coincidental. In practice, the peak positions would fluctuate and statistically align with the posterior probability distribution across multiple realizations.
  Beyond this, an application of the recovery of these parameters is the mass function inference of the MBHs, which characterizes the features of their host galaxies that are very hard to probe electromagnetically.  If we assume that the scaling of EMRI rate with MBH mass is known, the hyper-parameters can be directly converted to the slope index of MBH mass function. Then, TianQin will provide a unique window on the MBH mass function and serve as a key diagnostic for deriving the mechanism that forms black hole seeds. The corresponding parameter estimation accuracy is approximately at the current level of observational uncertainty of the MBH mass function.  In the loss cone model, the cusp regrowth time, which affects the galaxy as a nursery for EMRI formation, is related to redshift. By measuring  $\sigma_z$, we can learn more about the impact of redshift on the cusp regrowth time and gain a better understanding of galaxy evolution.

%During the exploration, we also found that the MBH spin will not influence the SNR very much, although a slight change of the MBH spin would greatly change the waveform shapes. This may be because all the EMRI waveforms are truncated at the Schwarzschild last stable orbit in this study, and the energy dissipated via GW has not much difference.  Consequently,  the selection bias for spin population models can be neglected.

\section{Conlusions}\label{conclusion}

In this study, we investigate the constraints on the EMRI population model with TianQin. We analyzed the EMRI population model using a parametrization method, where we use a broken power-law function to describe the MBH mass distribution, use a truncated Gaussion function labeled as model-1 and a power-law plus Gaussion function labeled as model-2 to describe the redshift distribution.  We utilized AK to calculate the EMRI waveform and truncated them at the Schwarzschild LSO and the Kerr LSO. Then, we used the SNR to estimate the number of EMRIs detectable by TianQin during its mission lifetime, and employed the Fisher information matrix to determine the posterior distribution of these detectable EMRI sources. After calculating the selection bias, we explored the posterior of the hyper-parameters using the hierarchical Bayesian inference method.

Our results show that TianQin could recover the posterior distribution of hyper-parameters  describing the EMRI population model relatively precisely for model-1. The inferred population distributions are generally consistent with the true population distributions for both AKS and AKK.  With more detectable EMRI sources, the estimation precision for those hyper-parameters will be improved, and the confidence intervals of the posterior distributions will be narrowed. In the case with AKK, the $\alpha_1, \alpha_2$, b could be measured with precision of $46.4\%$, $12.6\%$ and $3\%$, respectively. The hyper-parameters  $\mu_z$ and $\sigma_z$ could be measured with a precision of $15.4\%$ and $21.1\%$.  All the hyper-parameters can be recovered within $1\sigma$ confidence interval.  For model-2, the hyper-parameters $\lambda$ and $\alpha$, describing the redshift distribution, cannot be well recovered, especially for AKS.  In the case with AKK, the inferred MBH mass distribution and the redshift distribution are generally consistent with the true distribution, but with large uncertainty for $\alpha$. The result shows that $\alpha_1, \alpha_2$, b could be measured with precision of $45.0\%$, $12.7\%$ and $4\%$, respectively. In comparison, $\lambda$, $\alpha$, $\mu_z$ and $\sigma_z$ could be measured with precision of $28.2\%$, $147.5\%$, $16.8\%$, and $24.1\%$, respectively.

\section*{Acknowledgements}
We are grateful to Hui-Tong for his helpful discussion and
advice. This work has been supported by Hebei Natural Science Foundation with Grant No. A2023201041, Postdoctoral Fellowship Program of CPSF under Grant Number GZC20240366.
Jian-dong Zhang is supported by the Guangdong Basic and Applied Basic Research Foundation(Grant No. 2023A1515030116).
~\\

%%%%%%%%%%%%%%%%%%%%%%%%%%%%%%%%%%%%%%%%%%%%%%%%%%
%\section*{Data Availability}

%The underlying SkyMapper and ANU 2.3m/WiFeS data will be made available on reasonable request to the authors.



\bibliographystyle{unsrt}
\bibliography{reference} % if your bibtex file is called example.bib

\begin{thebibliography}{10}

\bibitem{Talbot:2017yur}
Colm Talbot and Eric Thrane.
\newblock {Determining the population properties of spinning black holes}.
\newblock {\em Phys. Rev. D}, 96(2):023012, 2017.

\bibitem{LIGOScientific:2018jsj}
B.~P. Abbott et~al.
\newblock {Binary Black Hole Population Properties Inferred from the First and
  Second Observing Runs of Advanced LIGO and Advanced Virgo}.
\newblock {\em Astrophys. J. Lett.}, 882(2):L24, 2019.

\bibitem{LIGOScientific:2020kqk}
R.~Abbott et~al.
\newblock {Population Properties of Compact Objects from the Second LIGO-Virgo
  Gravitational-Wave Transient Catalog}.
\newblock {\em Astrophys. J. Lett.}, 913(1):L7, 2021.

\bibitem{KAGRA:2021duu}
R.~Abbott et~al.
\newblock {Population of Merging Compact Binaries Inferred Using Gravitational
  Waves through GWTC-3}.
\newblock {\em Phys. Rev. X}, 13(1):011048, 2023.

\bibitem{LIGOScientific:2016vbw}
B.~P. Abbott et~al.
\newblock {GW150914: First results from the search for binary black hole
  coalescence with Advanced LIGO}.
\newblock {\em Phys. Rev. D}, 93(12):122003, 2016.

\bibitem{amaro2017laser}
Pau Amaro-Seoane, Heather Audley, Stanislav Babak, John Baker, Enrico Barausse,
  Peter Bender, Emanuele Berti, Pierre Binetruy, Michael Born, Daniele
  Bortoluzzi, et~al.
\newblock Laser interferometer space antenna.
\newblock {\em arXiv preprint arXiv:1702.00786}, 2017.

\bibitem{TianQin:2015yph}
Jun Luo et~al.
\newblock {TianQin: a space-borne gravitational wave detector}.
\newblock {\em Class. Quant. Grav.}, 33(3):035010, 2016.

\bibitem{Baker:2019nia}
John Baker et~al.
\newblock {The Laser Interferometer Space Antenna: Unveiling the Millihertz
  Gravitational Wave Sky}.
\newblock 7 2019.

\bibitem{TianQin:2020hid}
Jianwei Mei et~al.
\newblock {The TianQin project: current progress on science and technology}.
\newblock {\em PTEP}, 2021(5):05A107, 2021.

\bibitem{alexander2017emris}
Tal Alexander.
\newblock Emris and the relativistic loss-cone: The curious case of the
  fortunate coincidence.
\newblock In {\em Journal of Physics: Conference Series}, volume 840, page
  012019. IOP Publishing, 2017.

\bibitem{Pan:2021oob}
Zhen Pan, Zhenwei Lyu, and Huan Yang.
\newblock {Wet extreme mass ratio inspirals may be more common for spaceborne
  gravitational wave detection}.
\newblock {\em Phys. Rev. D}, 104(6):063007, 2021.

\bibitem{Naoz:2022rru}
Smadar Naoz, Sanaea~C. Rose, Erez Michaely, Denyz Melchor, Enrico Ramirez-Ruiz,
  Brenna Mockler, and Jeremy~D. Schnittman.
\newblock {The Combined Effects of Two-body Relaxation Processes and the
  Eccentric Kozai\textendash{}Lidov Mechanism on the Extreme-mass-ratio
  Inspirals Rate}.
\newblock {\em Astrophys. J. Lett.}, 927(1):L18, 2022.

\bibitem{Bortolas:2019sif}
Elisa Bortolas and Michela Mapelli.
\newblock {Can supernova kicks trigger EMRIs in the Galactic Centre?}
\newblock {\em Mon. Not. Roy. Astron. Soc.}, 485(2):2125--2138, 2019.

\bibitem{Fan:2022wio}
Hui-Min Fan, Shiyan Zhong, Zheng-Cheng Liang, Zheng Wu, Jian-dong Zhang, and
  Yi-Ming Hu.
\newblock {Extreme-mass-ratio burst detection with TianQin}.
\newblock {\em Phys. Rev. D}, 106(12):124028, 2022.

\bibitem{Ye:2023aeo}
Chang-Qing Ye, Jin-Hong Chen, Jian-dong Zhang, Hui-Min Fan, and Yi-Ming Hu.
\newblock {Observing white dwarf tidal stripping with TianQin gravitational
  wave observatory}.
\newblock {\em Mon. Not. Roy. Astron. Soc.}, 527(2):2756--2764, 2023.

\bibitem{Hu:2018yqb}
Xin-Chun Hu, Xiao-Hong Li, Yan Wang, Wen-Fan Feng, Ming-Yue Zhou, Yi-Ming Hu,
  Shou-Cun Hu, Jian-Wei Mei, and Cheng-Gang Shao.
\newblock {Fundamentals of the orbit and response for TianQin}.
\newblock {\em Class. Quant. Grav.}, 35(9):095008, 2018.

\bibitem{Brown:2020uvh}
Warren~R. Brown, Mukremin Kilic, A.~B\'edard, Alekzander Kosakowski, and
  P.~Bergeron.
\newblock {A 1201 s Orbital Period Detached Binary: the First Double Helium
  Core White Dwarf LISA Verification Binary}.
\newblock {\em Astrophys. J. Lett.}, 892(2):L35, 2020.

\bibitem{Kremer:2017xrg}
Kyle Kremer, Katelyn Breivik, Shane~L. Larson, and Vassiliki Kalogera.
\newblock {Accreting Double white dwarf binaries: Implications for LISA}.
\newblock {\em Astrophys. J.}, 846(2):95, 2017.

\bibitem{Korol:2017qcx}
Valeriya Korol, Elena~M. Rossi, Paul~J. Groot, Gijs Nelemans, Silvia Toonen,
  and Anthony G.~A. Brown.
\newblock {Prospects for detection of detached double white dwarf binaries with
  Gaia, LSST and LISA}.
\newblock {\em Mon. Not. Roy. Astron. Soc.}, 470(2):1894--1910, 2017.

\bibitem{Feng:2019wgq}
Wen-Fan Feng, Hai-Tian Wang, Xin-Chun Hu, Yi-Ming Hu, and Yan Wang.
\newblock {Preliminary study on parameter estimation accuracy of supermassive
  black hole binary inspirals for TianQin}.
\newblock {\em Phys. Rev. D}, 99(12):123002, 2019.

\bibitem{Ruan:2021fxq}
Wen-Hong Ruan, He~Wang, Chang Liu, and Zong-Kuan Guo.
\newblock {Rapid search for massive black hole binary coalescences using deep
  learning}.
\newblock {\em Phys. Lett. B}, 841:137904, 2023.

\bibitem{Shuman:2021ruh}
Kevin~J. Shuman and Neil~J. Cornish.
\newblock {Massive black hole binaries and where to find them with dual
  detector networks}.
\newblock {\em Phys. Rev. D}, 105(6):064055, 2022.

\bibitem{Katz:2019qlu}
Michael~L. Katz, Luke~Zoltan Kelley, Fani Dosopoulou, Samantha Berry, Laura
  Blecha, and Shane~L. Larson.
\newblock {Probing Massive Black Hole Binary Populations with LISA}.
\newblock {\em Mon. Not. Roy. Astron. Soc.}, 491(2):2301--2317, 2020.

\bibitem{Fragione:2022ams}
Giacomo Fragione and Abraham Loeb.
\newblock {Constraining the Cosmic Merger History of Intermediate-mass Black
  Holes with Gravitational Wave Detectors}.
\newblock {\em Astrophys. J.}, 944(1):81, 2023.

\bibitem{Torres-Orjuela:2023hfd}
Alejandro Torres-Orjuela, Shun-Jia Huang, Zheng-Cheng Liang, Shuai Liu,
  Hai-Tian Wang, Chang-Qing Ye, Yi-Ming Hu, and Jianwei Mei.
\newblock {Detection of astrophysical gravitational wave sources by TianQin and
  LISA}.
\newblock {\em Sci. China Phys. Mech. Astron.}, 67(5):259511, 2024.

\bibitem{Klein:2022rbf}
Antoine Klein et~al.
\newblock {The last three years: multiband gravitational-wave observations of
  stellar-mass binary black holes}.
\newblock 4 2022.

\bibitem{Buscicchio:2021dph}
Riccardo Buscicchio, Antoine Klein, Elinore Roebber, Christopher~J. Moore,
  Davide Gerosa, Eliot Finch, and Alberto Vecchio.
\newblock {Bayesian parameter estimation of stellar-mass black-hole binaries
  with LISA}.
\newblock {\em Phys. Rev. D}, 104(4):044065, 2021.

\bibitem{Ewing:2020brd}
Becca Ewing, Surabhi Sachdev, Ssohrab Borhanian, and B.~S. Sathyaprakash.
\newblock {Archival searches for stellar-mass binary black holes in LISA data}.
\newblock {\em Phys. Rev. D}, 103(2):023025, 2021.

\bibitem{Toubiana:2020cqv}
Alexandre Toubiana, Sylvain Marsat, Stanislav Babak, John Baker, and Tito
  Dal~Canton.
\newblock {Parameter estimation of stellar-mass black hole binaries with LISA}.
\newblock {\em Phys. Rev. D}, 102:124037, 2020.

\bibitem{Zhang:2022xuq}
Xue-Ting Zhang, Chris Messenger, Natalia Korsakova, Man~Leong Chan, Yi-Ming Hu,
  and Jing-dong Zhang.
\newblock {Detecting gravitational waves from extreme mass ratio inspirals
  using convolutional neural networks}.
\newblock {\em Phys. Rev. D}, 105(12):123027, 2022.

\bibitem{Wardell:2021fyy}
Barry Wardell, Adam Pound, Niels Warburton, Jeremy Miller, Leanne Durkan, and
  Alexandre Le~Tiec.
\newblock {Gravitational Waveforms for Compact Binaries from Second-Order
  Self-Force Theory}.
\newblock {\em Phys. Rev. Lett.}, 130(24):241402, 2023.

\bibitem{Lynch:2021ogr}
Philip Lynch, Maarten van~de Meent, and Niels Warburton.
\newblock {Eccentric self-forced inspirals into a rotating black hole}.
\newblock {\em Class. Quant. Grav.}, 39(14):145004, 2022.

\bibitem{Isoyama:2021jjd}
Soichiro Isoyama, Ryuichi Fujita, Alvin J.~K. Chua, Hiroyuki Nakano, Adam
  Pound, and Norichika Sago.
\newblock {Adiabatic Waveforms from Extreme-Mass-Ratio Inspirals: An Analytical
  Approach}.
\newblock {\em Phys. Rev. Lett.}, 128(23):231101, 2022.

\bibitem{Vazquez-Aceves:2022dgi}
Veronica Vazquez-Aceves, Yiren Lin, and Alejandro Torres-Orjuela.
\newblock {Sgr A* Spin and Mass Estimates through the Detection of an Extremely
  Large Mass-ratio Inspiral}.
\newblock {\em Astrophys. J.}, 952(2):139, 2023.

\bibitem{Renzini:2022alw}
Arianna~I. Renzini, Boris Goncharov, Alexander~C. Jenkins, and Pat~M. Meyers.
\newblock {Stochastic Gravitational-Wave Backgrounds: Current Detection Efforts
  and Future Prospects}.
\newblock {\em Galaxies}, 10(1):34, 2022.

\bibitem{LISACosmologyWorkingGroup:2022kbp}
Nicola Bartolo et~al.
\newblock {Probing anisotropies of the Stochastic Gravitational Wave Background
  with LISA}.
\newblock {\em JCAP}, 11:009, 2022.

\bibitem{Boileau:2020rpg}
Guillaume Boileau, Nelson Christensen, Renate Meyer, and Neil~J. Cornish.
\newblock {Spectral separation of the stochastic gravitational-wave background
  for LISA: Observing both cosmological and astrophysical backgrounds}.
\newblock {\em Phys. Rev. D}, 103(10):103529, 2021.

\bibitem{Amaro-Seoane:2007osp}
Pau Amaro-Seoane, Jonathan~R. Gair, Marc Freitag, M.~Coleman~Miller, Ilya
  Mandel, Curt~J. Cutler, and Stanislav Babak.
\newblock {Astrophysics, detection and science applications of intermediate-
  and extreme mass-ratio inspirals}.
\newblock {\em Class. Quant. Grav.}, 24:R113--R169, 2007.

\bibitem{Amaro-Seoane:2012lgq}
Pau Amaro-Seoane.
\newblock {Relativistic dynamics and extreme mass ratio inspirals}.
\newblock {\em Living Rev. Rel.}, 21(1):4, 2018.

\bibitem{Zi:2021pdp}
Tie-Guang Zi, Jian-Dong Zhang, Hui-Min Fan, Xue-Ting Zhang, Yi-Ming Hu, Changfu
  Shi, and Jianwei Mei.
\newblock {Science with the TianQin Observatory: Preliminary results on testing
  the no-hair theorem with extreme mass ratio inspirals}.
\newblock {\em Phys. Rev. D}, 104(6):064008, 2021.

\bibitem{Shi:2019hqa}
Changfu Shi, Jiahui Bao, Haitian Wang, Jian-dong Zhang, Yiming Hu, Alberto
  Sesana, Enrico Barausse, Jianwei Mei, and Jun Luo.
\newblock {Science with the TianQin observatory: Preliminary results on testing
  the no-hair theorem with ringdown signals}.
\newblock {\em Phys. Rev. D}, 100(4):044036, 2019.

\bibitem{Gair:2008bx}
Jonathan~R. Gair.
\newblock {Probing black holes at low redshift using LISA EMRI observations}.
\newblock {\em Class. Quant. Grav.}, 26:094034, 2009.

\bibitem{Gair:2010yu}
Jonathan~R. Gair, Christopher Tang, and Marta Volonteri.
\newblock {LISA extreme-mass-ratio inspiral events as probes of the black hole
  mass function}.
\newblock {\em Phys. Rev. D}, 81:104014, 2010.

\bibitem{Gair:2010bx}
Jonathan~R. Gair, Alberto Sesana, Emanuele Berti, and Marta Volonteri.
\newblock {Constraining properties of the black hole population using LISA}.
\newblock {\em Class. Quant. Grav.}, 28:094018, 2011.

\bibitem{Huang:2020rjf}
Shun-Jia Huang, Yi-Ming Hu, Valeriya Korol, Peng-Cheng Li, Zheng-Cheng Liang,
  Yang Lu, Hai-Tian Wang, Shenghua Yu, and Jianwei Mei.
\newblock {Science with the TianQin Observatory: Preliminary results on
  Galactic double white dwarf binaries}.
\newblock {\em Phys. Rev. D}, 102(6):063021, 2020.

\bibitem{Wang:2019ryf}
Hai-Tian Wang et~al.
\newblock {Science with the TianQin observatory: Preliminary results on massive
  black hole binaries}.
\newblock {\em Phys. Rev. D}, 100(4):043003, 2019.

\bibitem{Liu:2020eko}
Shuai Liu, Yi-Ming Hu, Jian-dong Zhang, and Jianwei Mei.
\newblock {Science with the TianQin observatory: Preliminary results on
  stellar-mass binary black holes}.
\newblock {\em Phys. Rev. D}, 101(10):103027, 2020.

\bibitem{Fan:2020zhy}
Hui-Min Fan, Yi-Ming Hu, Enrico Barausse, Alberto Sesana, Jian-dong Zhang,
  Xuefeng Zhang, Tie-Guang Zi, and Jianwei Mei.
\newblock {Science with the TianQin observatory: Preliminary result on
  extreme-mass-ratio inspirals}.
\newblock {\em Phys. Rev. D}, 102(6):063016, 2020.

\bibitem{Liang:2021bde}
Zheng-Cheng Liang, Yi-Ming Hu, Yun Jiang, Jun Cheng, Jian-dong Zhang, and
  Jianwei Mei.
\newblock {Science with the TianQin Observatory: Preliminary results on
  stochastic gravitational-wave background}.
\newblock {\em Phys. Rev. D}, 105(2):022001, 2022.

\bibitem{Zhu:2024qpp}
Liang-Gui Zhu, Hui-Min Fan, Xian Chen, Yi-Ming Hu, and Jian-dong Zhang.
\newblock {Improving Cosmological Constraints by Inferring the Formation
  Channel of Extreme-mass-ratio Inspirals}.
\newblock {\em Astrophys. J. Suppl.}, 273(2):24, 2024.

\bibitem{Babak:2017tow}
Stanislav Babak, Jonathan Gair, Alberto Sesana, Enrico Barausse, Carlos~F.
  Sopuerta, Christopher P.~L. Berry, Emanuele Berti, Pau Amaro-Seoane, Antoine
  Petiteau, and Antoine Klein.
\newblock {Science with the space-based interferometer LISA. V: Extreme
  mass-ratio inspirals}.
\newblock {\em Phys. Rev. D}, 95(10):103012, 2017.

\bibitem{Mandel:2018mve}
Ilya Mandel, Will~M. Farr, and Jonathan~R. Gair.
\newblock {Extracting distribution parameters from multiple uncertain
  observations with selection biases}.
\newblock {\em Mon. Not. Roy. Astron. Soc.}, 486(1):1086--1093, 2019.

\bibitem{Gair:2022fsj}
Jonathan~R. Gair, Andrea Antonelli, and Riccardo Barbieri.
\newblock {A Fisher matrix for gravitational-wave population inference}.
\newblock {\em Mon. Not. Roy. Astron. Soc.}, 519(2):2736--2753, 2022.

\bibitem{Thrane:2018qnx}
Eric Thrane and Colm Talbot.
\newblock {An introduction to Bayesian inference in gravitational-wave
  astronomy: parameter estimation, model selection, and hierarchical models}.
\newblock {\em Publ. Astron. Soc. Austral.}, 36:e010, 2019.
\newblock [Erratum: Publ.Astron.Soc.Austral. 37, e036 (2020)].

\bibitem{Pan:2021lyw}
Zhen Pan, Zhenwei Lyu, and Huan Yang.
\newblock {Mass-gap extreme mass ratio inspirals}.
\newblock {\em Phys. Rev. D}, 105(8):083005, 2022.

\bibitem{Ruan:2018tsw}
Wen-Hong Ruan, Zong-Kuan Guo, Rong-Gen Cai, and Yuan-Zhong Zhang.
\newblock {Taiji program: Gravitational-wave sources}.
\newblock {\em Int. J. Mod. Phys. A}, 35(17):2050075, 2020.

\bibitem{Adams:2012qw}
Matthew~R. Adams, Neil~J. Cornish, and Tyson~B. Littenberg.
\newblock {Astrophysical Model Selection in Gravitational Wave Astronomy}.
\newblock {\em Phys. Rev. D}, 86:124032, 2012.

\bibitem{Farr:2019rap}
Will~M. Farr.
\newblock {Accuracy Requirements for Empirically-Measured Selection Functions}.
\newblock {\em Research Notes of the AAS}, 3(5):66, 2019.

\bibitem{Tiwari:2017ndi}
Vaibhav Tiwari.
\newblock {Estimation of the Sensitive Volume for Gravitational-wave Source
  Populations Using Weighted Monte Carlo Integration}.
\newblock {\em Class. Quant. Grav.}, 35(14):145009, 2018.

\bibitem{Barack:2003fp}
Leor Barack and Curt Cutler.
\newblock {LISA capture sources: Approximate waveforms, signal-to-noise ratios,
  and parameter estimation accuracy}.
\newblock {\em Phys. Rev. D}, 69:082005, 2004.

\bibitem{Ye:2023lok}
Chang-Qing Ye, Hui-Min Fan, Alejandro Torres-Orjuela, Jian-dong Zhang, and
  Yi-Ming Hu.
\newblock {Identification of gravitational waves from extreme-mass-ratio
  inspirals}.
\newblock {\em Phys. Rev. D}, 109(12):124034, 2024.

\bibitem{Babak:2009ua}
Stanislav Babak, Jonathan~R. Gair, and Edward~K. Porter.
\newblock {An Algorithm for detection of extreme mass ratio inspirals in LISA
  data}.
\newblock {\em Class. Quant. Grav.}, 26:135004, 2009.

\bibitem{Chua:2021aah}
Alvin J.~K. Chua and Curt~J. Cutler.
\newblock {Nonlocal parameter degeneracy in the intrinsic space of
  gravitational-wave signals from extreme-mass-ratio inspirals}.
\newblock {\em Phys. Rev. D}, 106(12):124046, 2022.

\bibitem{Finn:1992wt}
Lee~S. Finn.
\newblock {Detection, measurement and gravitational radiation}.
\newblock {\em Phys. Rev. D}, 46:5236--5249, 1992.

\bibitem{Vallisneri:2007ev}
Michele Vallisneri.
\newblock {Use and abuse of the Fisher information matrix in the assessment of
  gravitational-wave parameter-estimation prospects}.
\newblock {\em Phys. Rev. D}, 77:042001, 2008.

\bibitem{Heinzel:2024hva}
Jack Heinzel, Matthew Mould, and Salvatore Vitale.
\newblock {Nonparametric analysis of correlations in the binary black hole
  population with LIGO-Virgo-KAGRA data}.
\newblock {\em Phys. Rev. D}, 111(6):L061305, 2025.

\bibitem{Thorne:1974ve}
Kip~S. Thorne.
\newblock {Disk accretion onto a black hole. 2. Evolution of the hole.}
\newblock {\em Astrophys. J.}, 191:507--520, 1974.

\bibitem{Chapman-Bird:2022tvu}
Christian E.~A. Chapman-Bird, Christopher P.~L. Berry, and Graham Woan.
\newblock {Rapid determination of LISA sensitivity to extreme mass ratio
  inspirals with machine learning}.
\newblock {\em Mon. Not. Roy. Astron. Soc.}, 522(4):6043--6054, 2023.

\bibitem{foreman2013emcee}
Daniel Foreman-Mackey, David~W Hogg, Dustin Lang, and Jonathan Goodman.
\newblock emcee: the mcmc hammer.
\newblock {\em Publications of the Astronomical Society of the Pacific},
  125(925):306, 2013.

\bibitem{Lyu:2023ctt}
Xiangyu Lyu, En-Kun Li, and Yi-Ming Hu.
\newblock {Parameter estimation of stellar mass binary black holes in the
  network of TianQin and LISA}.
\newblock {\em Phys. Rev. D}, 108(8):083023, 2023.

\end{thebibliography}

%\bibliographystyle{unsrtnat}
%\bibliography{beta2187}
%\begin{thebibliography}{99}
%\end{thebibliography}

%%%%%%%%%%%%%%%%%%%%%%%%%%%%%%%%%%%%%%%%%%%%%%%%%%

%%%%%%%%%%%%%%%%% APPENDICES %%%%%%%%%%%%%%%%%%%%%

%%%%%%%%%%%%%%%%%%%%%%%%%%%%%%%%%%%%%%%%%%%%%%%%%%

% Don't change these lines
%\bsp	% typesetting comment
%\label{lastpage}
\end{document}